# Magnetocatalytic Adiabatic Spin Torque Orbital Transformations for Novel Chemical and Catalytic Reaction Dynamics: The Little Effect

*Reginald B. Little

**Department of Chemistry**
**Florida A&M University**
**Tallahassee, Florida 32307**

**Abstract**
　　In this manuscript the theory and phenomena associated with the Little Effect are introduced as the spin induced orbital dynamics of confined fermions under strong magnetic and thermal environments.  This Little Effect is considered in details for the electron transfer reactions associated with redox processes of Cu-Ag alloy within deionized water and for the orbital dynamics during the iron catalyzed covalent bond rearrangements associated with amorphous carbon conversion to diamond.  Furthermore, prolong extreme conditions of 74,000 amps, 403 V, strong Lorentz compression, and thermal stresses upon this Cu-Ag- $H_2O$ system on the basis of the Little Effect of high spin, thermally induced orbital dynamics are predicted and demonstrated to cause the magnetically organized reverse beta, electron capture, proton capture and neutron capture processes for various infrequent pycnonuclear transmutations within the Cu-Ag coil.  The general experimental verification and the broad implications of this Little Effect on chemistry are demonstrated within these two ideal systems: an ionic case and a molecular case.  The Little Effect is contrasted with the Hedvall Effect as a dynamical phenomenon causing the kinematics of the Hedvall Effect.  The compatibility of the Little Effect with the Woodward-Hoffmann Rule is demonstrated.  The Little Effect provides greater understanding of order in systems far from equilibrium.  The implications of the Little Effect for other interesting phenomena such as ferromagnetism, unconventional magnetism, superparamagnetism, superconductivity, and pycnonuclear effects are concluded.

# Magnetocatalytic Adiabatic Spin Torque Orbital Transformations for Novel Chemical and Catalytic Reaction Dynamics: The Little Effect

*Reginald B. Little

**Department of Chemistry**
**Florida A&M University**
**Tallahassee, Florida 32307**

**Introduction**

   The Little Effect is phenomenon associated with the orbital (directional) dynamics and rehybridization induced by spin interactions within and between atoms and molecules and even within bulk materials. This effect manifests itself in chemical reactions, catalytic processes, conduction processes, photochemical processes, photophysical events and other processes. On the basis of spin induced orbital dynamics and rehybridization, the Little Effect recognizes intrinsic, important fleeting transient magnetic influences (ferromagnetic or superparamagnetic ordering) during these fermionic dynamics. Traditionally magnetic influences on chemical reactions have not been sufficiently included [1]. The much weaker strength of the magnetic force relative to its electric source has been the basis for this weak magnetic account. It is on this basis that chemical reactions have been understood by the electric properties of electron affinity, ionization energy, electronegativity, nucleophilicity and electrophilicity. *But here we demonstrate that the nonclassic approximation of such systems and the consequent Fermi-Dirac statistics and Pauli Rule determine important spin and magnetic factors during dynamic processes just as these spin factors determine the configurations and structures of atomic and molecular stationary states. Therefore for fermion dynamics and transformations, spin and magnetics modulate chemical reactions according to the Little Effect.* This work determines the fermionic nature of important intermediates during adiabatic processes within these systems: Cu-Ag-water and Fe-C. The influence of spin under strong external magnetization already been demonstrated to affect reaction kinetics via picosecond time resolved spectroscopy by Steiner [2]. The Little Effect determines that even fleeting multi-spin intermediary states contribute novel physicochemical dynamics. On this basis, the Little Effect determines magnetic factors of fermions, bosons, spin density, coercivity, permeability, spinphillicity, spin angular momenta and orbital angular momenta for understanding chemical, catalytic and pycnonuclear reaction dynamics. The Little Effect gives a twist to proton transfer, acid catalysis [3], and Lewis acidity, providing dynamical explanations to concerted proton transfer [4] in some tautomeric systems.

   To a greater extent the spin and magnetics have been considered during nonadiabatic photophysical and photochemical processes [5], basically on the premise that the stronger electric factors have been weakened in the photo-excited states. G. N. Lewis [6] first suggested that magnetism may be used to study radicals and their reactions. In these cases, radical pair effects [7] and phosphorescence [5] have been understood. The power of Pauli antisymmetry in slowing electronic relaxation under the electric potential has been thoroughly demonstrated during photochemical reactions and optical processes by radical pair effects and phosphorescence, respectively. Indeed, the power of Pauli antisymmetry is thought to holdup the stars against gravitational collapse and strong and weak interactions. Such power of antisymmetry is therefore nontrivial even within the electronics and dynamics within and between atoms. The Little Effect is a manifestation of this power of antisymmetry in affecting atomic and molecular physics. In some systems, photo-excitation has been discovered to induce intersystem crossing of electrons: El-Sayed Effect [8]. This El-Sayed Effect demonstrates the power of orbital motion for influencing spin during nonadiabatic processes. The Kasha Effect [9] demonstrates the slower rate of many optical transitions (spontaneous emission) relative to phonon relaxation. Unlike the spontaneous emission limited by Kasha's Effect, the Einstein Effect [10] involves the acceleration of optical transitions from excited states by the photons of similar energy and phase: stimulated emission. Here the Little Effect considers aspects of internal processes involving the adiabatic stepwise electronic transformation by spin organized orbital dynamics, orbitally induced intersystem crossing, efficient phonon effects and internally optically stimulated phenomena within dense fermionic media for novel explanations of ferromagnetism, superconductivity, paramagnetic catalysis, and pycnonuclear processes. On the basis of the radical pair effect, the ability of the nuclear spin to torque electron spin via orbital motion for isotopic radical pair effects has led to Buchachenko's magnetic isotope effect (MIE) [11]. On the basis of molecular systems with unsaturated atoms and sufficient through bond exchange, molecular magnetism [12] has been developed on the basis of electron-electron spin --- spin interactions. Also recently, the spin polarization of conduction electrons in some systems has led to the realization of spintronics [13]. In these semiconducting systems, the Dresselhaus Effect [14] and the Rashba Effect [15] have noted the orbitally induced spin transitions during conduction by voltage induction within the bulk and on the surface during conduction, respectively. For all these cases, the spin and magnetics have been limited to the application of the Pauli Principle for influencing changes in kinetics and dynamics of fermionic pairing or un-pairing during the chemical reactions, the electronic conduction, and/or the photophysical, photochemical processes. For such antisymmetric modulated processes, the spin polarization slows chemical bonding, electronic transport and/or optical transitions on the basis of Pauli's Principle. The El-Sayed, Dresselhaus and Rashba Effects do determine orbitally driven spin dynamics by photophysical and electric field conditions, nonadiabatically. ***In this work, however, the Little Effect is introduced as the accumulation of spin and the resulting alteration of orbital dynamics without photons but only by phonons and dense magnons alone. The accumulation of spin, the spin densification and the associated novel multi-spin induced electronic effects under adiabatical conditions have not been predicted, observed and reported until here.*** Buchachenko have reported the idea of spin catalysis [16] but for non-adiabatic photochemical systems, involving individual radical pairs. ***But the Little Effect is based on multi-spin interactions of dense radicals and fermionic medias and the corresponding ordering.*** The spin symmetry breakage by the intrinsic magnetic effects of spin and orbital motion associated with El-Sayed, Rashba and Dresselhaus Effects and the complimenting orbital symmetry breakage by spin and spin waves presented here by the Little Effect are consistent with the theory that magnetic field can cause dynamical symmetry breakage [17]. Order within systems far from equilibrium has been demonstrated [18] thermodynamically. But here the Little Effect determines the microscopic and the dynamical bases for such order in systems far from equilibrium. It is on this basis that Little Effect explains and determines the dynamics and the basis for order under extreme conditions that organize diamond formation, superconductivity and pycnonuclear reactions. ***We here introduce multi-spin catalytic effects during adiabatic processes. Here we introduce a new spin basis for explaining order within systems far from equilibrium. Here we introduce the novel effect of such unusual high spin environments under prevailing thermal conditions that create unique spin interactions for unusual electronic orbital transformations. For the first time, the Little Effect introduces the phenomena of multi-spin interactions inducing processes for changing orbital, motional dynamics during multi-fermion interactions for novel processes in high spin media. Furthermore, the Little Effect introduces important adiabatic (without external photon) spin driven dynamics for novel magnetic field effects during chemical and physical processes.***

On the basis of the Little Effect, catalytic properties of ferrometals are better understood. Such ferro-catalysts provide illustrative high spin environments for demonstrating novel multi-spin dynamics. The change in catalysis by ferro-metals was first observed in the 1930s as the change in the kinetics of catalytic processes by ferro-metal catalysts as the temperature is raised beyond the Curie temperature [19]. These phenomena have become known as the Hedvall Effect [20]. Examples of the Hedvall Effect include: the temperature dependent hydrogen reduction of $Fe_2O_3$ (hematite) to $Fe_3O_4$ (magnetite) [21]; the kinetics of oxidation of a silver-cadmium alloy [22]; and the reactions of silver sulfate and barium oxide [23]. For the reaction dynamics of hydrocarbons on and within Fe, Co, and Ni catalysts, the Hedvall Effect is consistent with " Mechanistic Aspects of Carbon Nanotube Nucleation and Growth" [24] on the basis of spin density phenomena due to temperature fluctuations about the Curie temperature and the consequent thermal and spin (phonon and magnon) impact on hydrocarbon decomposition, carbon uptake, hydrogen desorption, carbon diffusion and carbon release as graphitic nets on the surfaces of Fe, Co, and Ni nanoparticles. Siegel et al. [25] have computationally demonstrated the Hedvall Effect on the basis of this Comprehensive Mechanism by Little [24], such that ferromagnetism below the Curie temperature expels carbon from the interior of the catalysts to lower the free energy whereas temperature fluctuations above the Curie point cause carbon dissolution within the catalysts to lower the system's free energy. Prior to Siegal et al [25], qualitative descriptions of these effects were given by Little in The Comprehensive Mechanism of Carbon Nanotube Nucleation and Growth [24] and also in The Resolution of the Diamond Problem [26]. ***On the basis of the Little Effect, spin dynamics within and on the catalyst particles induce orbital dynamics of carbon for its rehybridization and intersystem crossing to modulation the decomposition, transport and formation of graphitic structures during the nucleation and growth of the***

*carbon nanotubes.* This is a special case of the Little Effect whereby the Hedvall Effect causes carbon nanotube formation in these ferromagnetic systems. The Hedvall Effect slows and/or accelerates catalytic decomposition, transport, and precipitation processes of carbonaceous states associated with phenomena of the Little Effect to synchronize carbon nanotube nucleation and growth for avoiding carbon accumulation and catalytic poisoning.

This basis of the Hedvall Effect involves the thermal effects on magnetics for kinematical influences on the magnetocatalysis. The Little Effect is more dynamical, whereas the Hedvall effect is more kinematical. However, the Little Effect involves the intrinsic spin and magnetic effects on the fermion dynamics during such magnteocatalysis for the transformations and the transport of the reactants and intermediates into products. The Little Effect explains the Hedvall Effect. Transient magnetic ordering in superparamagnetic systems is determined by the Little Effect. Furthermore, novel orbital dynamics driven by transient magnetic states for both ferromagnetic and paramagnetic systems under external, strong magnetic environments are embraced by the Little Effect. ***The Little Effect determines that strong magnetic environments frustrate the Hedvall Effect such that high spin transient states persist even under greater thermal energies that lead to novel spin induced orbital transformations for novel catalysis, chemical and even pycnonuclear reaction dynamics.*** Many have been misled by the high temperatures beyond the Curie temperature and its disordering of spins. It is on this basis that many investigators have limited magnetic field effects to ultrastrong magnetic fields wherein the magnetic field is dominating the thermodynamics over its electric source. ***However, it is important to note on the basis of the Little Effect that the dynamics of phonons are much slower than the electronic motion so that between collisions and vibrations substantial electronic spin ordering occurs under external magnetization, allowing important high spin interactions for consequent important spin induced orbital motional changes and novel chemical, conduction and catalytic phenomena. So even at high temperatures, strong external magnetic field can reorganize spins for novel dynamics before phonons can scatter.*** On this basis of the Little Effect, atoms in high spin, temperate environments exhibit unique chemical and catalytic reactions by orbital dynamics not just the simple kinetics and dynamics of re-pairing by Pauli antisymmetry. It is on this basis of the Little Effect that diamond was discovered to nucleate in the open atmosphere without high pressure and H plasma but by strong magnetization of Fe and amorphous carbon at 900 $^{\circ}$C [26, 27]. In this manuscript, the Fe-C system and a Cu-Ag-$H_2O$ system are presented for demonstrating the Little Effect and the novel chemistry and catalysis in high spin, high temperature environments. These two systems are ideal for demonstrating the Little Effect on the basis of Cu and Ag manifesting the dependence of redox processes on the slow kinetics of s-d rehybridization and carbon exhibiting kinetic difficulty for rehybridization to $sp^3$ for the tetrahedral assembly of diamond. The Little Effect involves thermally driven spin induced rehybridization to catalyze these reactions.

**Procedure**

These novel spin induced orbital dynamics for novel magneto-catalytic phenomena were studied within two magnetized systems: the water oxidation of copper and silver alloy metal and the iron catalyzed transformation of amorphous carbon to diamond. The Fe catalyzed transformation of amorphous carbon to diamond has been reported elsewhere [26, 27], but this system is considered more here to demonstrate the Little Effect in this manuscript. The water oxidation of Cu-Ag is extensively explored and developed here in this manuscript as more experimental evidence of the Little Effect for determining novel electrocatalysis, electrochemistry, electromagnetochemistry and pycnonuclear phenomena. In particular, the DC magnets at the National High Magnetic Field Laboratory (NHMFL) in Tallahassee, Florida were analyzed because such magnets operate by forcing huge electric currents through Cu-Ag coils by high volts to generate very strong magnetic fields. The Cu-Ag coils produce huge heat loads, which are removed by flowing large volumes of deionized water through and around the coils. In this work, these DC magnets were recognized as very unique environments to explore subtle magnetic field effects on chemical reactions due to the rapidly flowing, corroding water, Cu-Ag coils, strong magnetic field (up to 45 tesla), large electric field, Lorentz pressure stresses and thermal stresses. The DC magnets operate at 403 volts and 74,000 amps. The coils are stacked, separated by thin insulating polymer and compressed tightly within the magnet. The coils consist of over 954 kg (2103 lbs) of Cu-Ag metal alloy. More than 20,000 gallons per minute of highly deionized water flow through the holes in the coils to cool and maintain their temperature at 40-65 $^{\circ}$C during the operation of the magnet. During operation, the water corrodes, oxidizes and dissolves the Cu-Ag coil. Magnetic field effects on electrochemistry of corrosion and dissolution are explored here. The predicted low rates of pycnonuclear reactions on the basis of the Little Effect were also explored for hydrogen absorption within the Cu-Ag lattice. Water samples were collected from the magnet during its operation at various magnetic field strengths, coil temperatures and operation modes (ramping or stepping the field). The water samples were collected in acidic media to prevent the precipitation of the dissolved metals. The water was analyzed by inductively coupled plasma mass spectroscopy in order to correlate the relative amount of Cu and Ag solute dissolved under different magnetic field strengths. Isotopic analyses of the water samples were also done in order to measure $^2H/^1H$ and $^{18}O/^{16}O$ ratios. The Cu-Ag coils were analyzed by SIMS and Rutherford back scattering spectroscopy to measure relative amounts of $^2H$ and $^1H$ within the coils. The SIMS analyses also determine the unusual presence of other heavier nuclides in the coils.

**Results**

As the magnetic field was rapidly ramped from 11 tesla to 45 tesla, Cu and Ag levels increased dramatically in the cooling water. See Table 1. The Ag increased and the Cu decreased as the field is rapidly ramped from 11 tesla to 45 tesla at 12 tesla per minute with slight temperature increase from 8.9 C° to 42.1 C°. The rapid ramping of the magnetic field appears to have affected the Cu more than the Ag. However, as the field and mechanical vibrations equilibrated at 45 tesla for 151 second, the Cu level dropped but the Ag level remained high. This suggests that the ramping of the magnetic field caused mechanical shock that caused the large level of Cu erosion from the coils. The sustained high levels of Ag suggest that its increase is a true magnetic field effect. The Ag levels dropped when the magnetic field was rapidly dropped from 45 tesla to 11 tesla and the coil remained warm. The Cu level also dropped as the magnetic field was rapidly ramped down from 45 tesla to 11 tesla. The temperature during such ramping of the magnetic field changed only slightly so the increase in metal content of the water is attributed to the magnetic field effects or possibly mechanical effects and friction between the coils. In zero applied magnetic field, the Cu-Ag ratio varies from 10.53 to 4.57 as the temperature was raised from 22 °C to 100 °C. See Table 2. On the basis of these data, the low Cu-Ag ratio of about 1 at 45 tesla (42 °C) cannot be a result of only thermal effects. Another effect is contributing to the enhanced lowering of the Cu/Ag ratio. This additional effect is that due to the strong magnetic field. In order to discriminate the magnetic field effects from frictional and mechanical effects, the water was sampled while stepping the magnetic field by 3-4 tesla steps with mechanical and thermal equilibration before sampling the water. The step experiments revealed increase in Ag dissolution for stronger magnetic field. The data from these steps experiments also revealed relative differences in the Cu dissolution relative to the Ag dissolution. As the magnetic field strength increased beyond 22 tesla, the Cu/Ag ratio in the water decreased, dramatically. See Figures 1. The Cu/Ag ratio dropped from 2 at 11 tesla to a Cu/Ag ratio of 1 at 42.5 tesla. The isotopic analysis of the water revealed deuterium ($^2H$) enrichment and $^{18}O$ depletion of the water under stronger magnetic field. The $\delta^2H$ increased from -90+/- 8 at 11 tesla to -54 +/- 8 at 45 tesla. The $\delta^{18}O$ decreased from -1.5 +/- 0.2 at 11 tesla to -2.6 +/- 0.2 at 45 tesla. See Table 3. In Table 4, the SIMS data also determined higher levels of deuterium to protium in used Cu-Ag magnet coils relative to unused Cu-Ag coils. On the basis of SIMS analysis of the Cu-Ag coils before and after the $^1H/^2D$ decreased from 7324+/-100 to 5331+/-221, respectively. Sputtering the surface and depth analysis by SIMS yielded before and after use $^1H/^2D$ levels of 6881+/- 224 and 5619+/-122, respectively, within the interior of the Cu-Ag Coils. SIMS also revealed traces of tritium or helium-3 and helium-4 within the Cu-Ag coils and Fe used to form diamond in the magnetic field. SIMS analysis revealed higher levels of Pd, Ru, Fe, Ti and Rh after the Cu-Ag coils have been used relative to the levels of these rare nuclides in the initially unused Cu-Ag coils. ICP mass spectroscopy of the coils revealed unusual $^{103}Rh/^{105}Pd$ ratios and $^{99}Ru/^{103}Rh$ ratios for some of the used Cu-Ag coils. See Table 5. These ratios deviated from the relative natural isotopic abundances for these various nuclei. Moreover, these nuclides have been suspected by other prior investigators to be subject to unusual low temperature transmutations.

**Table 1 – ICP-MS of Cu and Ag Levels in Cooling Water While Rapidly Ramping the Magnetic Field**

| Time (s) | Magnetic Field (tesla) | Temperature (C⁰) | Cu | Ag |
|---|---|---|---|---|
| 0 | 11.5 | 8.9 | 15921 | 15707 |
| 265 | 45.2 | 29.5 | 39535 | 18849 |
| 416 | 45.2 | 42.1 | 13293 | 18844 |
| 596 | 11.5 | <42.1 | 7481 | 15147 |

| | | | | | |
|---|---|---|---|---|---|
| 776 | 11.5 | | 8.2 | 7882 | 14360 |

**Table 2 ICP-MS of Cu and Ag Levels in Cooling Water at Various Temperatures**

| Temp (°C) | time (min) | $^{63}Cu/^{65}Cu$ | $^{107}Ag/^{109}Ag$ | $^{63}Cu/^{107}Ag$ |
|---|---|---|---|---|
| 22 | 10 | 2.33 | 1.06 | 7.14 |
| 22 | 45 | 2.35 | 1.06 | 10.53 |
| 100 | 10 | 2.40 | 1.07 | 2.96 |
| 100 | 15 | 2.39 | 1.05 | 3.81 |
| 100 | 45 | 2.52 | 1.06 | 4.57 |
| Inlet water | | 2.36 | 1.06 | 15.54 |

**Table 3 - Isotopic Modifications Within the Cooling Water**

| Sample | $\delta^2H$ | $\delta^{18}O$ |
|---|---|---|
| Local water | -89+/- 8 | -1.8+/-0.2 |
| 11 tesla | -90+/-8 | -1.5+/-0.2 |
| 21.5 tesla | -42+/-8 | -2.5+/-0.2 |
| 45 tesla | -54+/-8 | -2.6+/-0.2 |

**Table 4 SIMS H/D Ratio in Cu-Ag Coils**

| Sample | Unused Cu-Ag H/D | Used Cu-Ag H/D | Used Cu-Ag with surface CuO and $Ag_2O$ H/D |
|---|---|---|---|
| Cs source | 7474 | 5069 | 6646 |
| | 7238 | 5152 | 6101 |
| | 7260 | 5707 | 6490 |
| | | 5397 | 6723 |
| Average | 7324 | 5331 | 6490 |
| Ave. Deviation | 100 | 221 | 195 |
| | | After Presputtering | |
| | 7034 | 5375 | |
| | 6920 | 5810 | |
| | 6432 | 5659 | |
| | 7136 | 5631 | |

|  |  |  |
|---|---|---|
| Average | 6881 | 5619 |
| Ave Deviation | 224 | 122 |

### Table 5 - Isotopic Modifications within the Cu-Ag Coils by ICP-MS and SIMS

| Sample | ICP-MS $^{103}$Rh/$^{105}$Pd | ICP-MS $^{99}$Ru/$^{103}$Rh | Comments from ICP-MS | Comments from SIMS |
|---|---|---|---|---|
| Cu-Ag Coil Before Use | 2.250 | 0.0042 | Ru is present | |
| Cu-Ag Coil After Use | 2.281 | 0.0038 | Ru is present | |
| Cu-Ag Coil After Use | 2.347 | 0.0027 | Rh is present | $^2$H, $^3$H, $^4$He, $^{103}$Rh, $^{105}$Pd |
| Cu-Ag Coil After Use | 2.223 | 0.0038 | Ru is present | |
| Cu-Ag Coil After Use | 2.257 | 0.0027 | Rh is present | $^2$H, $^3$H, $^4$He, $^{103}$Rh, $^{105}$Pd |
| Cu-Ag Coil After Use | 2.193 | 0.0028 | Rh, and Pd are present | $^2$H, $^3$H, $^4$He, $^{103}$Rh, $^{105}$Pd |
| Cu Coil After Use | 2.247 | 0.0034 | | |
| Resin Water | 2.250 | 0.0030 | Rh is present | |
| Control Values | 2.243 | 0.0034 | | |

**Discussion**

Within the strong DC magnets, redox reactions contribute to the corrosion and oxidation of the conductive coils that generate the strong magnetic field by the large current densities pushed by huge potentials. These reactions reduce some of the hydrogen and oxidize some of the oxygen of the cooling water. The resistive heating of these Cu-Ag coils requires continuous cooling by high flowing deionized water during the operation of the magnet. Such sporadic heating contributes interesting thermal effects on the coil erosion, oxidation-reduction and pycnonuclear reactions. See Tables 1 and 2. The outlet cooling water is continuously stripped of dissolved Cu and Ag ions and $O_2$ and recirculated in a closed loop back through the huge magnet. The coils experience huge compressive and tensile stresses due to their compression under bolt in zero applied field and the consequent Lorentz forces by the applied strong magnetic fields (45 tesla). Such Lorentz forces are substantial, and in some cases cause sufficient force for mechanical structural failure, thereby limiting the nature of the coil material on the basis of mechanical strength in the design and construction of these strong magnets. The below electron transfer reactions occur under these extreme novel conditions of strong magnetic field (up to 45 Tesla), large electric potentials (403 volts), large current densities (74,000 amps), large thermal stresses, large Lorentz pressure stresses on the coil, and the ultrapure corrosive water with high flow rates. In this work, *such an extreme unusual environment was selected to explore the novel magnetic influence on redox chemical reactions (reactions 1-12) (Table 6) and pycnonuclear reactions (reactions 13-17) ( Table 7) for demonstrating the Little Effect.* The redox reactions between the Cu-Ag coils and the ultrapure water involve some of the following reactions:

**Table 6 -  Oxidation And Reduction Reactions between the Cu-Ag Coil and Cooling Water**

1. $H_2O\ (l) \leftrightarrow H^+\ (aq) + OH^-\ (aq)$
2. $OH^-\ (aq) \leftrightarrow H^+\ (aq) + O^{2-}\ (aq)$
3. $Cu(s) + H^+\ (aq) \leftrightarrow Cu^+\ (aq) + H\ (metal)$
4. $Cu^+\ (s) + H^+\ (aq) \leftrightarrow Cu^{2+}\ (aq) + H\ (metal)$
5. $Ag(s) + H^+\ (aq) \leftrightarrow Ag^+\ (aq) + H\ (metal)$
6. $Ag^+\ (s) + H^+\ (aq) \leftrightarrow Ag^{2+}\ (aq) + H\ (metal)$
7. $2Cu^+\ (s) + O^{2-}\ (aq) \leftrightarrow Cu_2O(s)$
8. $Cu^{2+}\ (s) + O^{2-}\ (aq) \leftrightarrow CuO(s)$
9. $2Ag^+\ (s) + O^{2-}\ (aq) \leftrightarrow Ag_2O(s)$
10. $Ag^{2+}\ (s) + O^{2-}\ (aq) \leftrightarrow AgO(s)$
11. $2H_2O\ (l) \leftrightarrow O_2(aq) + 4e^- + 4H^+$

12. $\qquad$ $4OH^- (aq) \leftrightarrow O_2 (aq) + 2H_2O + 4e^-$ 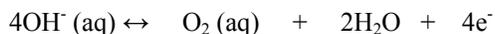

For these reactions, both Cu and Ag oxidations are endothermic. Within the strong electric potential (403 volts) and strong magnetic field (45 tesla) within the DC magnets, the electron transfer to water is hastened by the magneto-electrochemical contribution to the coil oxidation and dissolution, which drives the above reactions to the right. However, as the magnetic and electric fields are lowered to zero, the decrease in the magnetic field and the electric field and lower temperature cause the reactions (to some extent) to go back to the left. The observed large uptake of hydrogen (Tables 3 and 4 ; Figures 2 and 4) by the coils and high water velocity (20,000 gpm) through the (2103 lbs) coil contribute to substantial hydrogen and oxygen separation within the electric fields within the strong magnet for these significant redox processes across the magnet. The hydrogen uptake is substantial, because there are 954 kg (2103 lbs) of Cu and Ag making up the coil in the hybrid magnet for absorbing large amounts of hydrogen from reduction of the cooling water during operation of the magnet. The resulting metal hydrides and oxides exhibit very different properties relative to the pure metals: conductivity, redox chemistry, expansion volumes, magnetism, ect. Cu has a smaller free energy of oxidation than Ag's free energy of oxidation by the electron transfer to water. Obviously, differing effective nuclear charges of Cu and Ag contribute mostly to these differing free energies of oxidation. The activation energy for the oxidation of Cu by the water is also smaller than that for Ag. *The differing effective nuclear charges of Cu and Ag also contribute to the differing activation energies, but here we demonstrate that spin and magnetics contribute an additional although weaker factor (The Little Effect), which becomes more important in stronger external magnetic environments.* The larger free energy of Ag oxidation and its smaller electrical resistance has led to the use of this expensive metal for alloying and coating the Cu coils so as to supposedly extend the coil lifetime while maintaining the current density and heat removal. On the basis of this work, future high field magnets may require different coating than Ag due to the greater corrosion and dissolution of Ag into the cooling water under the effect of the stronger magnetic fields.

In this work on the basis of such kinetics and dynamics, *we predict and demonstrate that spin effects contribute to the electron transfer from the metal coil to the water*, such that spin alignment of the electrons and protons (within an external magnetic field) within the metal coil and the hydronium ions of the water (respectively) have higher activation energy for electron transfer from the metal to the solvated protons relative to spin unaligned electrons and protons of the metal coil and the hydronium ions (respectively) within the cooling water. Pauli antisymmetry of $e^-$ spins of the metal and the $p^+$ of the hydronium in the water slows $e^-$ transfer and oxidation-reduction reactions within the strong external magnetic field. Here on the basis of such spin and orbital momenta, it is interesting to consider how the electronic states of Cu and Ag influence their oxidation by the water. See Table 7. The $e^-$ transfer from d orbitals would exhibit different kinetics than $e^-$ transfer from sd hybrid and s orbitals of the metals. The $e^-$ transfer from the s orbital of the metal to the solvated $p^+$ occurs faster than the electron transfer from the d orbital of the metal to the solvated $p^+$. The rehybridization of s and d orbitals of the metal therefore influence the redox processes with the water. The rehybridization and orbital transitions of the metal may occur nonadiabatically or abiabatically. On the basis of the El-Sayed Effect [8], nonadiabatic orbital transitions from d to s may result in spin polarization of the excited $e^-$ in the upper level s orbital relative to the $e^-$ remaining in the lower level d orbital. *But here the Little Effect considers the absence of external photons and focuses on the adiabatic d to s transition and rehybridization caused only by phonons, spin and magnetic properties with no external photons.* The Little Effect reasons this by considering the Kasha Effect and the frequent, probable phonon events in conjunction with the unusual high spin environment and consequent superparamagnetism, paramagnetism and/or fleeting ferromagnetism within this system. On the basis of the Kasha Effect [9], phonons efficiently scatter electronic excited states. *The Little Effect determines that such efficient phonon interactions within this unusually large spin exchange environment would allow the reverse effect of phonon successively exciting orbital transitions via virtual states.* The Little Effect here considers the Kasha Effect [9] within high spin and strongly interacting spin systems. Such a high spin system would allow the phonon bath to classically (via virtual states) excite electrons (the reverse Kasha Effect) between (stationary) orbital states rather than the opposite efficient relaxation of excited orbital states by creating phonons within the bosonic paired electrons. *The Little Effect introduces the organization of heat by fermion motion by spin interactions for torque of orbital states for exciting virtual states with the consequent antisymmetry of the aligned fermion spins within the virtual states preventing their relaxation from these intermediary virtual states by the release of phonons. So that the adiabatic multi- phonon excitation of upper level high spin organized virtual orbital states is stabilized so*

*that further stepwise multi-phonon excitation can yield a high spin organized stationary upper level orbital state for the overall adiabatic stepwise spin induced orbital excitation. The Little Effect thereby determines an adiabatic spin, induced orbital excitation of fermions.* In this way, the Little Effect explains how order can arise in extreme conditions far from equilibrium. On the basis of the Little Effect, the Cu and Ag may by thermal activation and spin interactions excite $e^-$ from d to s orbitals. The higher spin state of Cu for the d to s orbital state contributes stabilization by the surrounding higher spin environment. The resulting $e^-$ in the excited s orbital has greater probability of being near the metal nucleus where it can undergo spin interactions with the nucleus by Buchachenko Effect [11] for intersystem crossing. The electron by such intersystem crossing can then more readily be transferred to the solvated proton of the water.

Within the earth's weak (background) magnetic field, the spins of electrons and protons are randomly oriented so that the electron transfer shows less spin effects. On the basis of G. N. Lewis [6], here, *a strong magnetic field is used to alter spin symmetry and to study the consequent change in kinetics and dynamics of this electron transfer redox reaction under a strong external magnetic field.* For the zero applied magnetic field, the Cu by Russell-Saunders coupling is better able to adiabatically spin torque electrons from its d into s orbitals (by the Little Effect) for faster nuclear coupled spin-orbitally induced intersystem crossing within the s excited orbital and faster electron transfer from the s orbital of Cu to solvated $p^+$ in the water for faster $Cu^{2+}$ formation relative to $Ag^+$ formation. This ability of Cu contributes to its smaller activation barrier for electron transfer to the water relative to the Ag. This ability of Cu also contributes to Cu's larger oxidation reaction rate constant relative to Ag's oxidation kinetics by the cooling water. These thermodynamic and kinetic differences of Cu and Ag redox chemistry result from electronic structural differences between Cu and Ag and their cations, according to Table 7:

**Table 7 - Electronic Configurations of Various Oxidation States**

$Cu \quad (4s^1 3d^2 3d^2 3d^2 3d^2 3d^2 \leftrightarrow 4s^1 3d^2 3d^2 3d^2 3d^2 3d^2)$
$Cu^+ \quad (4s^0 3d^2 3d^2 3d^2 3d^2 3d^2 \leftrightarrow 4s^1 3d^2 3d^2 3d^2 3d^2 3d^1)$
$Cu^{2+} (4s^0 3d^2 3d^2 3d^2 3d^2 3d^1 \leftrightarrow 4s^1 3d^2 3d^2 3d^2 3d^1 3d^1)$
$Ag \quad (5s^2 4d^2 4d^2 4d^2 4d^2 4d^1 \leftrightarrow 5s^1 4d^2 4d^2 4d^2 4d^2 4d^2)$
$Ag^+ \quad (5s^0 4d^2 4d^2 4d^2 4d^2 4d^2 \leftrightarrow 5s^1 4d^2 4d^2 4d^2 4d^2 4d^1)$
$Ag^{2+} (5s^0 4d^2 4d^2 4d^2 4d^2 4d^1 \leftrightarrow 5s^1 4d^2 4d^2 4d^2 4d^1 4d^1)$

Such electronic structural differences lead to different s-d rehybridization dynamics and the consequent intrinsic spin-magnetic influences on the orbital dynamics during the $e^-$ transfer and redox reactions. Thereby, Ag (under jj coupling for weaker spin effects) is less able to spin torque electrons from d into s excited orbitals (by the Little Effect) and thereby less able to intersystem cross its electrons for electron transfer to the water in zero applied field and lower temperatures. This weaker spin torque of Ag for electron transfer is consistent with the needed acid catalysis for many of Ag's redox and precipitation reactions. In such acidic medias, protons assist the torque of electrons between orbitals of Ag for catalytic redox processes. Similar proton catalysis via spin effects explains the distinct chemistry of the crystallization of $CaCO_3$ in water and heavy water in magnetic field as observed by Lundsen Madsen [3]. *The Little Effect is here providing a more thorough explanation of acid catalysis on the basis of intrinsic spin and magnetics of protons for catalyzing orbital dynamics in the catalyzed reactant atoms. Furthermore, the Little Effect is here integrating this proton catalytic effect to the observed magnetic field effect during the oxidation of Cu and Ag by the cooling water within the strong DC magnet. The considerations of redox processes and acid catalyzed processes here demonstrate the strong correlation of the proton/hydrogen with the d electrons of Ag for consequent electronic and chemical dynamics. Later this strong correlation is linked to pycnonuclear processes on the basis of the Little Effect.*
So in zero applied magnetic field, the electronic structural differences between Cu and Ag lead to greater spin torque rehybridization and redox electron transfer from Cu to form $Cu^{2+}$, than from Ag to form $Ag^+$. This explains the data at low magnetic field strength and temperature of Figure 1 and tables 1 and 2. However in strong enough external magnetic field (sufficient to saturate the permeability of the medias), the spins of the electrons within the metal coil and the protons of hydronium ions of the cooling water become polarized so as to influence the activation energy for electron transfer and slow the kinetics of the redox reactions by Pauli antisymmetry. Initially, this spin orientation by low external magnetic field contributes to larger free energy of activation for the electron transfer (from 11 tesla to 22 tesla). So initially, at lower magnetization (<22 tesla), the external magnetic field slows oxidation and dissolution of the metal coil. However, as the external magnetic field is intensified, the high spin density waves and phonons alters the dynamics of electron transfer on the basis of the Little Effect. The resulting phase change and its effect on electron transfer dynamics occur near 22 tesla in this Cu-Ag-$H_2O$ system. See Figure 1. On the one hand, on the basis of antisymmetry, the external magnetism raises the free energy of the redox process by orienting electrons of metal and the solvated protons within the water, causing the initial slight drop in Cu and Ag levels at lower fields (< 22 tesla). **But on the other hand, on the basis of the Little Effect, the stronger external magnetization (>22 telsa) and consequent higher spin polarization within the metal coil and water allow more spin---spin interactions for novel spin induced d to s orbital dynamics for enhanced nuclear coupled spin orbital induced intersystem crossing and e$^-$ transfer from Ag to the water.** Furthermore, as will be discussed subsequently, on this basis of the Little Effect, under sufficiently strong magnetization (even for short times) and thermal activation, the multi-spin interactions may torque electrons into protons and Ag nuclei for reverse beta processes and electron or proton capture processes for neutron formation within the core s sub-shells of Ag and within its nucleus in addition to the hydrogen atom formation demonstrated here during the redox chemistry of the Cu-Ag-water system.

This d to s orbital dynamics of electrons within the metals is driven by orbital and spin electron---electron interactions. **Both spin and orbital interactions with phonons can torque the electrons between orbitals, but the Little Effect determines that polarized high spin electronic systems provide stronger torque of the electrons between orbital states relative to only diamagnetic interactions and environments, having only orbital---orbital interactions.** This added contribution of spin determines the observed difference between Cu and Ag redox dynamics in varying magnetic fields. Neither the hydrogen nor its proton has multi electrons for such multi electron-electron interactions to torque the electronic orbital dynamics. **The proton can torque electrons between orbitals by proton --- electron orbital and spin interactions by the Little Effect for explaining acid catalysis and concerted double proton transfer processes [4]; however the pH in the magnet is close to neutral. However, the metals have more electrons and heavier nuclei than the hydrogen and oxygen of the water so such spin induced orbital effects and intersystem crossing within the metal's s orbital occur faster within the metal lattice relative to the water.** The strong magnetic field enhances the effect of spin induced d to s orbital electronic motion and dynamics and rehybridization must therefore be generated within the Cu-Ag metal in the magnets. It is important to note the effect of stronger electron---electron interactions in the metals. Such importance of metals for catalyzing hydrogen redox reactions is consistent with catalytic effects important for platinum –hydrogen electrodes in forming hydrogen gas [28]. Electron ---electron interactions within the metals cause the d to s orbital dynamics and within the resulting excited s orbital intersystem crossing occurs faster due to stronger s orbital electronic coupling to the nuclear. These predicted dynamics on the basis of the Little Effect are consistent with the observations of Podgorny et al [29] that desorbed hydrogen from metal hydrides exhibits different chemical reactivity than non absorbed hydrogen atoms. The intersystem crossing in the s orbital can under appropriate conditions also drive electrons into protons for reverse beta processes. Cu atoms (of the first d series) exhibits stronger intrinsic spin torque of electrons between d and s orbitals for stronger electron localization and magnetism in comparison to Ag atoms (which is of the second d series). However, Ag atoms exhibit greater delocalization and bonding than Cu atoms. **At higher temperatures Ag-Ag bonds can be broken by phonons, the resulting Ag radical localized states, interact more strongly than Cu's localized states such that the phonons and magnons in the Ag, resulting in localized states at higher temperatures in stronger magnetic field cause greater d to s orbital transformations in Ag according to the Little Effect with greater consequent oxidation of the Ag relative to Cu.** See Figure 1 for thermal and magnetic effects. See Table 2 for thermal effects alone. This prediction of the Little Effect is demonstrated in the data. See Figure 1. As the magnetic field is increased and temperature is increased the Cu/Ag ratio drops from 2 to 1, demonstrating the enhanced redox activity of Ag relative to Cu. **This phenomenon of the Little Effect therefore also accounts for anomalous properties of hydrogen in late second series transition metals.** These late 4d transition metals are more subject to exchange of electrons for forming metal bands, rather than binding hydrogen. But thermal activation can contribute to the breakage of M-M and allow M-H bonds to form for these second series transition metal, for oxidation of metal atoms and hydrogen uptake by the late second series transition metal. Such absorbed hydrogen can exist in these late 4d metals in various states of localization and delocalization. **On the basis of the Little Effect, here it is suggested that spin and magnetics of electrons and absorbed protons provide better explanations of recent anomalous effects of hydrogen in these late 4d transition metals [29].** These effects are modulated by temperature and manifest the Hedvall Effect, whereby the kinetics of the chemical reaction change by changes in the thermal conditions and the resulting changes in magnetic phase, which cause spin changes. But now beyond the Hedvall Effect, we demonstrate more intrinsic dynamics and catalysis by the higher spin, higher temperature environment, by the multi-spin interactions and by the consequent multi-spin torque for changing orbital motion and direction (Little Effect) within the stronger external magnetic field for demonstrating the Little Effect. In particular, this spin polarization and slowing of electron transfer by the Hedvall Effect occur for both Cu and Ag. However, intrinsic dynamical differences are demonstrated here by the observed greater slowing of Cu electron transfer and oxidation in comparison to that of Ag as the magnetic field and the temperature are increased. ***Whereas the slowing of the oxidation is a manifestation of the Hedvall Effect, the different change in the oxidation kinetics of the two metals in stronger magnetic environments are aspects of the Little Effect.*** Here it is suggested that such magnetic effects on 4d Pd and Ag in sufficiently strong magnetic fields and thermal activation can accumulate enough electric potential energy by Pauli antisymmetry for consequent electric field driven spin torque of electrons into protons within the metal lattice.

The Little Effect accounts for the differing electron transfer from Cu and Ag under the spin polarizing conditions by strong external magnetization on the basis of the polarized electrons within the metal inducing orbital rehybridization between s and d orbitals of the metals. In zero field and low temperature, the Cu due to Russell Saunders coupling allows stronger spin interactions (relative to Ag) to torque its electrons between 3d and 4s orbitals for easier rehybridization. The rehybridization is more driven by the intrinsic spin interactions within Cu relative to Ag at lower temperatures and lower external magnetic field strengths. The rehybridization and the consequent greater s character of the d electrons contribute to greater coulomb integral and spin polarization in Cu relative to Ag. These factors by the Little Effect therefore also contribute explanations to ferromagnetism. Without external magnetization, such intrinsic spin induced rehybridization (Little Effect) of d electronic orbitals into s electronic orbitals also allows greater spin-orbital coupling for intersystem crossing to unalign the electrons of Cu (with solvated protons) for faster electron transfer to the protons of the hydronium ions (relative to Ag). ***This intrinsic spin induced torque of electronic orbital motion and rehybridization in Cu contributes to greater localization of spin on Cu atoms and the greater magnetic moment of Cu relative to Ag atoms. On this basis, the Little Effect explains superparamagnetism, ferromagnetism and ferrimagnetism. On this basis, the Little Effect explains why 3d transition metals form more molecular magnets.*** The stronger external magnetization and slight temperature increase of the Cu and Ag coil increase the ordering of electron spins for stronger spin torquing electrons between s and d orbitals and even greater oxidation of the Ag, possibly to $Ag^{2+}$. In the strong magnetic field, the water may therefore become more acidic due to increased oxidation of the coils and greater charge of dissolved ions. However, the external magnetic field causes greater change in the ordering of spins in (the intrinsically weaker spin interacting) Ag than in the intrinsically stronger spin interacting Cu, so that the spin induced torque of electrons of Ag is much more enhanced by the external magnetic ordering in comparison to that for Cu. Ag coils are more permeable to the magnetic field relative to Cu. The external magnetization lowers the barrier for redox electron transfer from Ag by the higher polarized spin environment causing faster spin induced d to s electronic rehybridization (Little Effect) of Ag atoms in the metal relative to the Cu atoms. ***The consequent higher spin density and higher temperature cause phonon-magnons to better torque electrons into different orbitals (Little Effect). It is quite interesting that electron transfer and heat transfer is occurring simultaneously. The explanation given here by the Little Effect also gives explanation to superconductivity. Superconduction involves electron transfer between orbitals of different orbital momenta. On the basis of BCS theory, low temperature superconductivity involves delocalized Cooper pair with reversible phonon scattering. The Little Effect determines high temperature superconductivity can arise by higher energy phonons reversibly scattering conduction correlated fermions between orbitals of different spin symmetry. On the basis of the El-Sayed Effect and the Dresselhaus and Rashba Effects internal exchange of photons and electric potential may drive the current causing motion between orbitals to change spin polarity to generate fermions during phonon scattering of bosonic states. However the resulting high spin and magnetic interactions assisted by phonons maintain fermion pairing and rehybridize the high spins back into superconducting bosonic pairs on the basis of the Little Effect. Therefore, this effect of magnons-phonons for causing orbital transitions in high spin environments gives a basis for better understanding high temperature superconductivity.*** Furthermore, Ag having a heavier nucleus and larger atomic number has stronger spin-orbital coupling with an electron in an s orbital than Cu. Therefore this Little Effect of spin induced orbital rehybridization (within a strong magnetic environment) from d to s orbitals and the consequent nuclear spin-orbital coupled intersystem crossing within the s-d orbital hybrid in high spin, strong magnetic, higher temperature environments is greater for Ag than Cu. It is on this basis that Ag is observed in the data to exhibit faster oxidation kinetics than Cu at higher temperatures and within stronger magnetic fields. The Little Effect explains the greater frustration by Ag of the Hedvall Effect in a stronger magnetic environment at elevated temperature. Ag is more subject to the Little Effect than Cu in undergoing oxidation by water in stronger magnetic field and at higher temperatures. It is on this basis that the Cu/Ag ratio decreased from 2 to 1 in the cooling water as the magnetic field is elevated from 11 tesla to 45 tesla and the coils got slightly hotter. See Figure 1. In addition to this demonstration of the Little Effect for accounting for the different redox kinetics and dynamics of the water with Cu and Ag, these data demonstrate a contribution of spin orbital coupling and intersystem crossing to adiabatic chemical reaction dynamics. In addition to the faster redox processes of Ag in stronger magnetic field observed here, the Little Effect predicts more reproducible pycnonuclear reactions within Pd and Ag under strong magnetization and spin torque orbital effects of electrons into protons within the plasma of the metal lattices.

The observed isotopic effects on the reduction of hydrogen in the water and the oxidation of the Cu and Ag in the strong magnetic environment further substantiate the importance of underlying spin and magnetics for affecting the Cu-Ag-water redox chemistry. See Table 3 and Table 4. Here it is important to validate the spin contribution to the redox chemistry in the strong magnetic field, because thermal effects also play a role. Table 2 demonstrates thermal effects. However, Figure 1 determines additional magnetic effects beyond the thermal effects associated with the oxidation of the Cu-Ag coil. During the operation of the magnet in strong magnetic field, the water was observed to become enriched in deuterium ($^2H$) and to become depleted in $^{18}O$. See Table 3. This contrary isotope effect for the two heavier isotopes of hydrogen and oxygen within the water demonstrates that this is not a simple vapor-liquid phase transformation effect based on the prevailing higher thermal conditions. Vapor-liquid phase effects would lead to similar enrichment of both the heavier isotopes ($^2H$ and $^{18}O$) within the water in the slightly warmer (42 $^\circ$C) coils by the operation generating stronger magnetic environment due to faster vaporization processes of water molecules containing the lighter isotopes $^1H$ and $^{16}O$. If evaporation were the only cause of the isotope effects then both $^2H$ and $^{18}O$ would be affected in the same way by being enriched within the water by the warmer operating Cu-Ag coil caused by resistive heating during its operation. However, this mass-only induced isotopic enrichment of $^2H$ and $^{18}O$ by vapor-liquid phase changes by the cooling water was not observed. The observed enrichment (See Table 3) of the heavier deuterium relative to protium in the water and depletion of the heavier $^{18}O$ relative to $^{16}O$ demonstrate that a factor other than mass is differentiating the heavier isotopes in a contrary manner. This other factor is the differing spin properties of the $^1H$ and $^2H$ isotopes relative to the similar spin properties of $^{16}O$ and $^{18}O$. ***The contrary isotopic effects of $^2H$ enrichment and $^{18}O$ depletion of the cooling water during the magnet operation express that magnetic and spin phase transformation effects are relevant in addition to thermal effects and vapor-liquid processes.*** The mass differences of the H and O isotopes are important for isotopic fractionation by vapor-liquid effects. However, spin effects are also important for the intrinsic isotopic spin differences of $^1H$ and $^2H$ for different redox chemistry in the stronger magnetic field. On the basis of the spintronics, the statistics for $^1H$ and $^2H$ isotopes differ from the statistics for $^{16}O$ and $^{18}O$ isotopes in that $^1H$ and $^2H$ are fermionic and bosonic, whereas the $^{16}O$ and $^{18}O$ are both bosonic. ***The $^1H$ and $^2H$ (fermion and boson) isotopes exhibit different statistics so would behave differently in the magnetized environment in comparison to the weak earthly magnetic environment for different chemical kinetics of the redox reactions with the Cu-Ag coils in the strong magnetic field.*** The bosonic deuterium does not benefit from the same spin induced orbital motion for its reduction by electron transfer from the metal as the fermionic protium. So the lighter high spin protium isotope of hydrogen obtains smaller activation energy for reduction by the metal in the stronger magnetic environment than the heavier bosonic deuterium. See Reactions 1-6. So deuterium is less readily reduced and absorbed by the metal coils for slower depletion from the water relative to protium during the operation of the magnet, explaining the magneto-chemical deuterium enrichment in the cooling water under strong magnetization.

The catalytic activity of the consequent protium absorbed in the coil in activating reactions 7-12 is also differentiated from deuterium's catalysis on the basis of it being fermionic and deuterium ($^2H$) being bosonic. Such distinct protium and deuterium catalyses for oxide decomposition are in analog to acid catalytic precipitation of AgCl from aqueous solutions of $Ag^+$ and $Cl^-$. In order for the $Ag_2O$ to precipitate, the protium assists by spin, orbital and coulomb interactions. The oxides have been detected on the surface of Cu-Ag coils by x-ray electron spectroscopy. The distinct crystallization of $CaCO_3$ in ordinary and heavy water by Madsen Lundsen [3] under strong magnetic field is another example of this. ***It is important here to generalize this spin catalytic phenomenon on the basis of the Little Effect of explaining acid catalysis by spin induced orbital dynamics by hydrogen and/or protons. It is further important to recognize the general basis of the Little Effect as to how external magnetic field can organize and enhance such effects of acid catalysis as in this Cu-Ag-water system and also in the H-atom catalytic formation of diamond [26,27].*** Protium affords such spin and magnetic contributions to the catalysis, but deuterium lacks such spin and magnetic induced catalysis.

The absorbed protium in the coil during the operation of the magnet leads to enhanced proton spin catalyzed reactions in the coil under strong magnetization. ***These proton spin effects within the coil are explained by the Little Effect and expanded for explanations of diamond formation and most acid catalyzed reactions in chemistry.*** Here we consider such generalization for hydrogen atom spin catalyzing transformation of carbon to diamond and this second case of proton-acid spin catalyzed precipitation of Cu and Ag oxides. The importance of spin and Fermi-Dirac statistics of protium in comparison to Bose-Einstein statistics of deuterium leads to spin isotopic effects and isotopic fractionation during the redox chemistry cited here and also isotopic fractionation during diamond formation [30,31]. The $^{16}O$ and $^{18}O$ are both bosons and the processes associated with their formation of CuO and $Ag_2O$ are not directly sensitive to magnetic field. However $^{18}O^{2-}$ is more slowly oxidized to $OH^-$ and $O_2$ than $^{16}O^{2-}$ due to mass differences. $^{18}O^{2-}$ precipitates faster on the metal coils relative to $^{16}O^{2-}$, which is faster oxidized to $O_2$ for its enrichment in the outlet water. As the magnet is turned off the higher levels of protium observed in the coil are reduced but deuterium remains in the coils. This is further consistent with the SIMS observation of unusually high levels of deuterium within the magnet coils after prolong use in the magnets and the accumulation of CuO and $Ag_2O$ (reactions 7-10) on the coil surface. See Table 4. This SIMS detection of $^1H$ and $^2H$ in the coils is a direct consequence of reactions 3-6. During the prolonged operation of the magnets, the Cu-Ag coil undergoes oxidation by reducing hydrogen in the water and absorbing the reduced hydrogen. The hydrogen accumulation in the coils is consistent with their observed embrittled after use. Some of the $O^{2-}$ (mostly $^{18}O$) is deposited as metal oxide and some of the $O^{2-}$ (mostly $^{16}O$) is oxidized to $O_2$ (g). The observed black oxide formation on the coils is consistent with metal oxide formation. The formation of $O_2$ (g) is consistent with the release of $O_2$ gas over the cooling water storage vessel and with the enhanced lifetime of the coils by use of nitrogen blanket over the stored cooling water reservoir, which continuously strips the recycled water of this forming $O_2(aq)$. Such catalytic redox processes in strong magnetic field will result in new technology for hydrogen storage and fuel cells and future fusion reactors.

In addition to this demonstrated novel electrochemistry, the extremely high current (74,000 amps), high potential (403V), strong magnetic field (45 tesla), huge Lorentz pressure stress and thermal stress provide a very conducive environment for exploring the debated and controversial cold fusion. In this work, it is important to note the much greater electrode potentials (403 volts) on the Cu-Ag coil, much higher electric current (74,000 amps) through the electrodes and the much larger mass (2103 lbs) of the electrode in comparison to the much lower power and massive electrolytic systems of prior investigators exploring such low temperature pycnonuclear fusion. ***Unlike the prior attempts of other investigators [32] at exploring cold fusion, this Cu-Ag system used in this current work within the strong DC magnet further sustains these extreme current densities over much longer times (>2000 hours) with the added effects of very strong magnetization ( 45 tesla). The Cu-Ag coils studied in this work have been subjected to such extremely high currents, strong magnetic field, ultrapure deionized corrosive water, large thermal stress and huge Lorentz pressure stresses for over 2000 hours. None of the prior researchers have come close to these prolong extreme conditions as used in this current work.*** On this basis, new and very interesting data were predicted and observed. Although it would be more conducive to study Pd under these extreme conditions, the Cu-Ag coil currently used in the DC magnet was more readily available. The Ag-H exhibits to a lesser extent some of the anomalous properties as the palladium-hydrogen electrochemical system. Cu contributes a 3d character in this Cu-Ag system in conjunction to the debated Fe-H lattice for geothermal cold fusion in the earth's interior [33]. The Cu alloy with Ag yields some useful 3d character to the metal matrix for novel spin-magnetics for enhancing pycnonuclear phenemona according to the Little Effect. Here in this work, under these extreme conditions these very interesting pycnonuclear phenomena are observed.

***This more ideal experimental environment for exploring cold fusion is deeply grounded in the prior theories of magnetic field effects on nuclear processes within stellar systems. Although the Little Effect introduces spin torque effect for altering $e^-$ and $p^+$ orbital dynamics for nuclear processes at much more moderate magnetic fields strengths in the lattices of the transition metals, many investigators have reasoned substantial magnetic contribution to the thermodynamics of fusion reactions by ultrastrong magnetic fields ($>10^{12}$ tesla) in stellar bodies.*** Many researchers have suggested and determined, magnetic field effects on pycnonuclear reactions occurring within strongly correlated dense charged plasma. In 1986, Khersonskii [34] explored the catalysis of nuclear reactions of hydrogen in strong magnetic field. In 1990, Romonov [35] determined reactive effects of nuclear reactions controlled by magnetic field action on ferromagnetic materials. In 1995, Sekershitskii [36] considered the effect of strong magnetic field on energy yield of pycnonuclear reactions. In 2002, Romodanov [37] determined the tritium generation in a glow discharge on hydrogen

isotopes in magnetic field. In 2004, Kondratyev [38] determined enhanced neutron capture reactions in strong magnetic fields of magnetars. These various examples involve a range of conditions of magnetic field strength, kinetic energy and potential energy factors.

The magnetic fields and temperatures within the metal lattices (45-1000 tesla) considered here certainly do not approach stellar conditions (>$10^{12}$ tesla) at least not over macroscopic spatial dimensions. Certainly no terrestrial conditions may involve magnetic fields comparable to magnetars and neutron stars over large volumes or huge bulk forces as by the gravitational potentials as in stars, nor the thermal conditions of stars for large-scale uncontrolled fusions. For these stellar systems, the thermodynamics of fusion is determined by the high temperatures and ultrastrong magnetic fields (> $10^{12}$ tesla) and tremendous gravity. The metal lattice is much different from these stellar bodies and the thermodynamics and kinetics of cool nuclear reactions in these metal lattices are limited to different more local electrochemical phenomena. The electrochemical effects between dense $e^-$ and $p^+$ and metal nuclei cause pycnonuclear reactions at the lower kinetic energy densities relative to the much higher energy densities within the neutron stars and magnetars. On this basis, the metal lattice is more limited in its rates of catalyzed cool nuclear reactions. ***It is on this basis that this work poses no danger for explosive or dangerous uncontrolled low temperature fusion. This low temperature pycnonuclear process within the metal lattice may therefore be self regulating and a possible way to harness the energy of fusion controllably.*** However, the hydrogen in the Cu-Ag matrix of this system under the huge current density and electric stress, Lorentz pressure stress and thermal stress presents a different environment from the stars for the novel exploration of these less probable low temperature nuclear reactions.

The differing conditions in the metal lattice relative to the stars lead to different mechanisms of pycnonuclear processes within the metals lattice of lower probability. ***Such low frequency pycnonuclear reactions and the magnetic field enhancement are here predicted, explored and demonstrated here on the basis of the Little Effect, whereby the magnetic organization of dense electrons, protons and metal nuclei within the metal lattice and thermal and pressure fluctuation cause oscillations between delocalized and localized electronic and protonic states, involving the already demonstrated s-d rehybridization and nuclear coupled intersystem crossing characteristic of Cu and Ag lattice under prevailing magnetic and thermal conditions such that $e^-$ and $p^+$ become localized into s orbitals of the Ag with the strong electric field of the Ag nuclei strongly disrupting gamma exchange between the $e^-$ and $p^+$ and $e^-$ —Ag nuclei for electron capture by the proton (reverse beta) and/or $e^-$ and/or $p^+$ capture by Ag nuclei.*** Unlike the stars, the electric forces driving the pycnonuclear reactions in some of the metal lattices are more short range, involving fewer atoms and fermions than the longer range huge gravity in the stars, which is the potential energy basis for stellar fusion. Also in the metal lattices, the amount of hydrogen is diluted. ***These conditions tend to eliminate the possibility of explosive chain events for the pycnonuclear processes (and also contribute to difficult reproducibility) within the metal lattice relative to thermonuclear processes within stellar bodies and fusible explosives. This work, therefore, in no way implies stellar conditions within the metal lattice nor the possibility of dangerous explosives by these systems.*** This work however explains low temperature fusion by rare conditions within metal lattices that result in rare infrequent fusion of extremely low probabilities and rates under some difficult conditions of activation. ***On the basis of this current work, such infrequency may be the reason for the irreproducibility, the impracticality and the prevailing controversy surrounding these cold fusion phenomena [39]. However, this infrequency is an advantage that may lead to future discoveries of enhancing these slow rates controllably for better practical conditions for future safer, peaceful, environmentally friendly uses of fusion within such metal lattices for peaceful energy sources to benefit mankind in contrast to the more dangerous thermonuclear fusion weapons and systems. Here we demonstrate that the strong magnetization of the metal lattice may cause higher reproducibility and frequency for future practical beneficial uses.*** This first successful observation of pycnonuclear reactions in this work is suggested to result from the strong magnetization, large current density, and the long time period that these extreme conditions are applied to the Cu-Ag coil.

As already demonstrated in the current work, the Cu-Ag lattice is favorable plasma containing and accumulating electrons and protons and undergoing rehybridization and orbital dynamics for novel electrochemistry and redox reactions. The $e^-$ and $p^+$ are fermions in the metal lattice, and are therefore subject a strong applied magnetic field. These electronic and protonic orbital dynamics are closely coupled to the electron and proton motions and nuclear motions and interactions such that these lattice protons and electrons can slam into Ag and Pd nuclei for electron and proton capture processes or for the $e^-$, $p^+$ pair to inelastically barrel into each other, forming neutrons by the reverse beta process under the huge current density and strong magnetic perturbation in this Cu-Ag coil. Having spin and orbital momenta and magnetic moments, these $e^-$, $p^+$, and metal nuclei are organized by external magnetic field such that these pycnonuclear reactions are enhanced by the strong external magnetization. The magnetic field may contribute to faster more reproducible effects, of such pycnonuclear reactions. The Cu-Ag coil in the hybrid magnet is a very interesting system for analyzing the redox electrochemistry and low frequency pycnonuclear reactions under prevailing strong magnetization. So how do we overcome the infrequency of cold fusion? In the current work, the Cu-Ag lattice provides a favorable environment wherein such pycnonuclear reactions may be enhanced for greater frequency on the basis of the strong external magnetization. *Although the magnetic fields within the metal lattice are in comparable to the magnetic fields in neutron stars, the weaker magnetic fields here allow different mechanisms and processes of lower frequency for cold fusion by chemical-catalytic processes.* The ultrastrong magnetic fields in neutron stars allow symmetry breakage for gravity to compress large volumes of hydrogen into neutrons and for the consequent huge energy release and power. *In the case here with the metal lattice, the strong electric force of the metal nucleus provides the muscle, but the less frequent, rare hydrogen compression in the metal lattice is limited to individual hydrogen atoms in the form of hydride ($H^-$) species. Unlike gravitational compression involving many nuclei at the same time or thermonuclear fusion involving many high speed nuclei, the pycnonuclear processes involve individual localized hydrogenous species for lower frequencies and rare occurrences. The external magnetization organizes these individual rare events for increasing rates.* Hence the pycnonuclear reactions within the metal lattice are sporadic, self limiting and rare. The external magnetic field (11-45 tesla) can locally organize (steer) the dynamics for higher fusion rates in a controllable manner. *If the magnetization and enhanced pycnonuclear reaction led to even greater magnetization for even greater rates, then the process would be synergistically explosive. But such synergy of pycnonuclear reations and magnetization does not occur. The magnetization accelerates the pycnonuclear reactions but the pycnonuclear reactions do not enhance the magnetization so no synergism occurs. The magnetization is external to the pycnonuclear process so there is no concern for reaction run away. The process is therefore a controllable technology for tapping fusion energy*.

The mechanism based on magnetic orchestration of pycnonuclear reactions involves the following steps: 1.) under the prevailing conditions hydrogen uptake by the metal lattice and the high current density allow the formation of some amount of a hydride species ($H^-$); 2.) the thermal and pressure fluctuations and magnetization cause the electronic rehybridization of the background Cu-Ag lattice with consequent sporadic localization and delocalization of these electrons and protons of hydride species ($H^-$) within the Cu-Ag lattice; 3.) these protons and electrons of this hydride species exist delocalized in the 4d-like orbitals of the Cu-Ag lattice; 4.) localization of protons and electrons produces this hydride species in the metal lattice by the rehybridization of 3d, 4d, 4s, and 5s orbitals of the metal lattice; 5.) such localization by lattice rehybridization and confinement of $H^-$ within sd hybrid orbitals contribute to greater s character of the interacting electrons and protons in the form of ($e_a^- p^+ e_b^-$) or (hydride species) within the sd hybrid orbitals within the metal lattice; 6.) within the sd hybrid orbitals the ($e_a^- p^+ e_b^-$) with its net negative charge is strongly attracted in the localization to the nucleus ($M^{47+}$) of the metal atoms within the lattice; 7.) the ($e_a^- p^+ e_b^-$) is heavier and more classical in its interactions with the nucleus; 8.) as the ($e_a^- p^+ e_b^-$) approaches the nucleus the $e_a^-$ is driven into tighter orbital correlation with the $p^+$ in order to shield the proton from the nearby nucleus ($M^{47+}$) in this confined s orbital state for the local metal nuclear compression of the $e_a^-$ and $p^+$; 9.) the spin and magnetic properties of the confined ($e_a^- p^+ e_b^-$) state are more paramagnetic, an external magnetic field can therefore orients the nuclear spin of the metal atoms with the spin and orbital moments of the ($e_a^- p^+ e_b^-$); 10.) as the ($e_a^- p^+ e_b^-$) approaches the nucleus ($M^{47+}$), the nuclear spin torques the $e_b^-$ by nuclear spin-orbit interactions for its intersystem crossing, so $e_b^-$ changes correlation with the ($e_a^- - p^+$), thereby driving the $e_a^-$ into the $p^+$ for even tighter orbits, this orbital compression is strengthened by the huge nearby electric field of the metal nucleus within the s orbital of the metal atom; 11.) the resulting aligned spins of the metal nucleus ($M^{47+}$) and the $e_b^-$ organize the steering of $e_a^-$ into collapse onto the $p^+$ for reverse beta to form neutrons, $e_b^-$ may also collapse onto the metal nucleus; the $p^+$ may collapse onto metal nucleus; the resulting neutron may also collapse on the nucleus for various rare transmutation processes. See Table 7. 12.) the proximity (less than 0.5 Angstroms) of the $e_a^-$ --- $p^+$ to the $e_b^-$ and the metal nucleus ($M^{47+}$) within the s orbital allows huge local magnetic fields within the s orbital for extremely strong spin torque of $e_a^-$ into the $p^+$ thereby preventing gamma exchange as in isolated hydrogen thereby allowing the $e_a^-$ --- $p^+$ to form a neutron. It is within the s orbital with finite nonzero probability of the $e_a^-$ --- $p^+$ and $e_b^-$ having very close proximity to the metal nucleus that length scales of $<10^{-14}$ m are very small compared to larger atom size dimensions of $>10^{-10}$ m such that the magnetic forces within the s orbital are on the order of $1/(10^{-5})^2$ times the magnetic forces between lattice electrons in the domain of say a ferrometal. The magnetic forces between lattice electrons in the domain of a ferrometal of Fe are on the order of 1000 tesla. So the magnetic forces between the $e^-$ and $p^+$ and the metal nucleus for very close nuclear approach of the hydride species to the nucleus of a metal atoms is on the order of $10^{10}$ X 1000 tesla or $10^{13}$ tesla. Therefore within the s orbital of the metal lattice, the $e^-$ and $p^+$ of the hydride species would locally experience tremendous magnetic fields on the order of the magnetic fields in magnetars. An external magnetic field organizes (as in this work) the ($e_a^- p^+ e_b^-$) and metal nuclei for more favorable weak interactions, leading to enhanced cross-sections for fusion events. *In zero applied magnetic field, the proper spin and orbital orientations for such fusion processes are much more random and less likely. The important of such left-right symmetry during weak processes has been demonstrated by Yang and Lee [40]. Yang and Lee determined that within an external magnetic field, the nuclear spin oriented such that during the beta process the release of electron has specific momentum relative to the nucleus that released it. Here on the basis of the Little Effect, it is demonstrated that an external magnetic field can orient the e and nucleus for the reverse process of reverse beta for greater probability of such rare fusion events. The external magnetic field in this way organizes the spins for such symmetry for the reverse beta process and $e^-$ or $p^+$ capture process by the metal nucleus for greater rates and reproducibility of the pycnonuclear reactions. Without the external magnetization, the cross-section and probability are much lower.* Here these still slow nuclear processes within the strong magnetic environment, high current densities, Lorentz compression and thermal fluctuations are observed due to the long period of these conditions, more than 2000 hours. Although, the rates of pycnonuclear reactions are still very slow under the conditions within the strong magnet, even greater energy input via laser irradiation of the Cu-Ag matrix can promote much greater pycnonuclear fusion rates for future practical energy sources. Large magnetic field can build up huge potential energy due to Pauli antisymmetry with faster spin torque of electrons into protons for faster neutron formation (reverse beta processes) and neutron, electron and proton captures by Ag and Pd nuclei. The greater spin torque on orbital motion and the greater nuclear induced intersystem crossing also contribute more pycnonuclear phenomena in 4d relative to 3d transition metals in strong magnetic fields.

**Table 8 – Pycnonuclear Reactions**

| | |
|---|---|
| 13. | $p^+ + e^- \rightarrow n$ |
| 14. | $p^+ + n \rightarrow d^+$ |
| 15. | $d^+ + n \rightarrow t^+$ |
| 16. | $t^+ \rightarrow {}^3He^+ + e^-$ |
| 17. | ${}^3He + n \rightarrow {}^4He$ |

The data from this work give evidence for these pycnonuclear reactions occurring and this theory of the Little Effect of pycnonuclear phenomena within the Ag-H-Cu lattice under the sporadic thermal spikes, high current density, huge Lorentz compression, and strong magnetization. On the basis of the Little Effect and the above theory of magnetically organized pycnonuclear reactions, the accumulated protium within the Cu-Ag coil during the operation of the magnet at high magnetic field should lead to low frequency transmutations by reverse beta processes of the protiums and lattice electrons into neutrons and various $e^-$, $p^+$ and neutron capture processes by lattice nuclides. Such reverse beta processes within the Cu-Ag matrix during the operation of the magnet under the strong magnetic field are consistent with the already demonstrated novel electrochemistry of the Cu-Ag-$H_2O$ system in the strong magnetic field (45 tesla), huge current (74,000 amps), large Lorentz compression of the coils, and thermal stress. From the prior discussion, this unusual electrochemistry causes protium reduction and protium accumulation within the Cu-Ag coils during operation with $e^-$ loss (oxidation) by the Cu-Ag matrix for $Cu^{2+}$ and $Ag^+$ dissolution. The $e^-$ transfer from the Cu and Ag atoms to the solvated proton of the cooling water leads to absorbed $h^+$ and $d^+$ into the Cu-Ag lattice. The resulting lattice $h^+$ and $d^+$ fluctuate between delocalized $e^-$, $h^+$ and $d^+$ states and localized ($h^+,e^-$) and ($d^+,e^-$) pair states, respectively. The localized states of hydrogen correspond to $e^-$, $p^+$ confined within s orbitals, d orbitals and s-d hybrid orbitals within the Cu-Ag metal lattice. The delocalized states correspond to $e^-$ and $p^+$ wave functions over the d bands. The data here determine that such localized-delocalized fluctuating hydrogen states within the Cu/Ag lattice in the strong magnetic field cause the isotopic separation of protium and deuterium with the accumulation of protium in the Cu-Ag coil during operation and accumulation of deuterium in the cooling water under the extreme operating conditions of the magnet. *The resulting high concentrations of protium in the Cu-Ag matrix under the strong magnetic field, huge current density and thermal-pressure stresses result in some of*

*the protium undergoing reverse beta processes with lattice electrons during the localized-delocalized transitions, which are driven by the thermal fluctuations of the Cu-Ag matrix and the consequent spin induced orbital s-d rehybridization on the basis of the Little Effect, causing the $e^-, p^+$ to become localized in s orbitals with the consequent metal nuclei inducing intersystem crossing of the $e^-, p^+$ spins and the consequent Coulombically compression of the $e^-$ into the $p^+$ for reverse beta processes, neutron capture processes by the metal nuclei, $e^-$ capture processes by metal nuclei and/or $p^+$ capture processes by the metal nuclei.* These thermal neutrons from the reverse beta processes would be absorbed by surrounding protium, Cu, and Ag, producing neutron rich isotopes like deuterium, tritium, $^{50}$Ti, $^{57}$Fe, $^{65}$Cu and $^{109}$Ag. The observed thermal neutron absorption of the various nuclides correlates with tabulated standard thermal neutron capture cross-sections. These heavier nuclei are observed by SIMS. See Figures 2-5.

*Tritium and helium were detected in the used Cu-Ag coils whereas none was observed in the unused coils.* See Figures 2 and 4. These tritium and helium signals are real. These masses of 3 Da and 4 Da in the SIMS are not associated with clustering of hydrogen atoms. Clustering is not a factor because the observed H/D ratios are in the range of the natural relative abundances of protium (H) and deuterium (D) in the coils before and after use in the magnet. The terrestrial natural relative abundance of H and D determine a ratio of H/D of 6666. The H/D ratios demonstrate that the D is real, not a hydrogen cluster. Also at the higher masses, in Figure 4, the $^7$Li/$^6$Li ratio is also consistent with the natural relative abundance of lithium. Not only does the H not cluster with itself, it does not cluster with the $^{19}$F to form masses of 20 Da and 21 Da in Figures 2 and 4. If there were any tendency to cluster during SIMS, the $^{19}$F nuclide would do it with H. $^{19}$F is the most reactive nuclide, yet H$^{19}$F is not observed at masses 20 Da and 21 Da in Figure 2. The observed correspondences of SIMS peaks with the natural abundances of various isotopes eliminate the possibility that the signals for D, T, He and Li nuclides are due to H clusters. This is strong evidence that the tritium and helium peaks are not cluster artifacts. These peaks reflect the SIMS observing actual nuclides not clusters of H atoms. Clustering does occur during the SIMS but not for these nuclides of such lower masses. The clustering becomes noticeable in the masses greater than that of $^{63}$Cu, which is a major component of the metal matrix. There is an accumulation of H and D in the coil after prolonged use. This D accumulation is consistent with the formation of neutrons in the Cu-Ag coils and their absorption by the accumulated protium during the prolong operation of the magnet. The fusion of hydrogen in Pd during electrolysis has been debated [32, 33, 39]. Here it is demonstrated that such fusion can occur by reverse beta formation of neutron and neutron absorption by the lattice hydrogen. These lightest nuclides have a proclivity to absorb neutrons and undergo fusion transmutations. The D also absorbs the neutrons formed in the lattice to form tritium. Tritium transmutes to $^3$He. $^3$He can absorb neutrons to form $^4$He. See Table 8.

*Furthermore the nuclides of Ti in the Cu-Ag coil also were observed to exhibit unusual changes in SIMS before and after use of the coils.* See Figures 2 and 4. $^{48}$Ti and $^{47}$Ti appear in anomalous ratios. The mass intensity of $^{47}$Ti is too high and the mass intensity of $^{48}$Ti is too low. The mass intensity of $^{50}$Ti is too high in Figure 4. The mass intensity of $^{51}$V is too high. The mass intensity of $^{50}$V is too low. These Ti and V isotopes do not appear in the coil (after it's prolonged use) according to their natural relative abundances. These discrepancies in the natural abundances of Ti and V isotopes are not artifacts of clustering of lighter elements. The observed anomalous mass intensities of Ti and V are incompatible with clustering of lighter elements on the basis of the natural abundances of the lighter elements. In addition, the other elements in the vicinity of Ti and V do not exhibit anomalous relative mass intensities of their isotopes. The other elements of the first transition metal series show no unusual mass spectra in comparison to Ti, V, Fe, Cr, Mn and Ni. See Figures 2 and 4. On the basis of the relative abundance of the Ti and V, Ti's and V's incompatibility with possible clustering of lighter nuclides on the basis of their relative abundances; the observed absence of clustering of lighter nuclides; the observation of neighboring isotopes of elements in the region of Ti and V according to their natural relative abundances; and the unique structural properties of Ti and V, these anomalous Ti and V mass intensities are attributed to transmutations occurring in the Cu-Ag coil during the prolong operation of the magnet. Ti has been shown to cause anomalous low temperature fusion tendencies of hydrogen isotopes. Titanium metal has been demonstrated to extend the lifetime of tritium absorbed within the Ti lattice. Titanium's anomalous mass spectra are consistent with the Little Effect on the basis that the Ti and V atoms undergoing unique rapid d-s rehybridization relative to other transition metals based on titanium's and vanadium's unique electronic structures. These s-d orbital dynamics in Ti and V, subject the Ti and V nuclei to $e^-$ capture and neutron capture processes due to the $e^-$ and $p^+$ localized within Ti's and V's s orbitals during the rapid localization-delocalization dynamics of lattice $e^-$ and $p^+$ associated with the rapid s-d rehybridization

processes of Ti and V. The other elements of the first transition series (Sc, Mn, and Zn) lack these electronic structural factors for such high spin effects and orbital rehybridization for spin induced s-d rehybridization (Fe, Co, Ni) and/or rapid s-d orbital rehybridization (Ti and V) for enhancing the pycnonuclear reactions of reverse beta and $e^-$ or $p^+$ captures by the nuclides. Ti and V may be transmuting between each other. It is also possible that $^{107}$Ag is fissioning into $^{57}$Fe and $^{50}$Ti. The mass intensity $^{57}$Fe is too high. The mass intensity of $^{50}$Ti is too high. $^{109}$Ag is fragmenting into $^{58}$Ni and $^{51}$V. The mass intensity of $^{51}$V is too high. $^{58}$Ni exhibits an unusually high mass intensity relative to $^{60}$Ni.

***The nuclides of the ferromagnetic metals of the first d series also exhibit unusual mass spectral changes within coils before and after their prolong use in the magnet.*** See Figures 2 and 4. These anomalies in the relative abundances of the Fe and Ni nuclides are real and not artifacts due to clustering of lighter elements. The clustering of lighter elements has already been observed to be unnoticed. In addition, the observed SIMS intensities of the Fe and Ni nuclides are incompatible with the natural relative abundances of possible clusters of lighter nuclides on the basis of their natural relative abundances. Furthermore, the other elements in the vicinity of these ferro nuclides exhibit their natural relative abundances in their SIMS intensities. In Figure 4, the $^{57}$Fe and $^{58}$Ni nuclides are present in greater than the relative abundances with reference to $^{56}$Fe and $^{60}$Ni, respectively. $^{57}$Fe is formed by transmutation, clustering does not account for the large mass intensity of $^{57}$Fe because this mass is lower than the mass of the major components of the Cu and Ag matrix. The SIMS determines that clustering is more prevalent for masses heavier than the Cu nuclides. $^{58}$Ni may be forming the $^{57}$Fe by transmutations. It is also possible that $^{107}$Ag is fissioning into $^{57}$Fe and $^{50}$Ti. The mass intensity of $^{57}$Fe is too high. The mass intensity of $^{50}$Ti is too high. $^{109}$Ag may also be fragmenting into $^{58}$Ni and $^{51}$V. The mass intensity of $^{51}$V is too high. $^{58}$Ni exhibits unusual high mass intensity relative to $^{60}$Ni. In addition, SIMS of a Fe-C-H matrix used to nucleate diamond in strong magnetic field revealed even higher levels of $^3$H, $^3$He and $^4$He. On the basis of the mechanism of the pycnonuclear reactions these high spin nuclides are expected to exhibit more low temperature nuclear reactions. The greater proclivity for pycnonuclear reactions by these ferro nuclides follows from their stronger spin torque of orbital motion of $e^-$ and $p^+$ of hydride species into nuclei within their s orbitals these nuclides undergo faster transmutations thereby explaining their anomalous SIMS data among used and unused Cu-Ag coils.

After Cu, clustering becomes noticeable in the SIMS. Prior to Cu the clustering is not noticed in the SIMS on the basis of the correspondences between SIMS intensities and the natural relative abundances of many isotopes. See Figure 4. The SIMS intensities for isotopes of $^6$Li and $^7$Li appear consistently with their natural relative abundances (7.5 : 92.5). The SIMS intensities for isotopes of $^{24}$Mg $^{25}$Mg and $^{26}$Mg appear consistently with their natural relative abundances (78.99 :10.00 : 11.01). The SIMS intensities for isotopes of $^{28}$Si, $^{29}$Si and $^{30}$Si appear consistently with their natural relative abundances ( 92.23: 4.67: 3.10). Although the $^{28}$Si nuclide has a higher relative mass intensity than the $^{29}$Si and $^{30}$Si, clustering cannot account for the relative SIMS peak intensities for Si isotopes. The SIMS intensities for isotopes of $^{52}$Cr, $^{53}$Cr, and $^{54}$Cr appear consistently with their natural relative abundances (83.7: 9.5: 2.3). The SIMS intensities of $^{67}$Zn and $^{68}$Zn appear consistently with their natural relative abundances (4.1:18.8). However, the SIMS intensities of $^{58}$Ni and $^{60}$Ni appear inconsistently in accord with their natural relative abundances (68.1:26.2). The mass intensity of $^{58}$Ni is too high. The SIMS intensities of $^{63}$Cu and $^{65}$Cu are not present according to their natural relative abundances (69.17:30.83). There appears to be higher levels of $^{63}$Cu for a $^{63}$Cu/$^{65}$Cu ratio of over 4, rather than the natural $^{63}$Cu/$^{65}$Cu ratio of 2.6. Prior to Cu, the presence of these various elements according to their natural relative abundances gives strong evidence against these various peaks being artifacts of clustering of lighter elements. Therefore these signals for the masses less than 63 Da are due to single nuclides and not clusters. A few light nuclides ($^{28}$Si, $^{28}$Na, $^{37}$Cl and $^{72}$Ge) exhibit higher than relative amounts, but this is here associated with these elements forming from the fission of heavy Ag nuclides. The data matches the theory perfectly. All nuclides exhibiting unusual intensities can be paired with one another such that the mass of the pair is a mass of one of the Ag nuclides: $^{107}$Ag or $^{109}$Ag. All the other nuclides (occurring by natural relative abundances) are observed in the coil after use in the magnet for various reasons. The DC magnet is a huge system containing many parts of various compositions. The important point is that here these elements come from these various parts of the hybrid magnet according to their natural relative abundances and are not relevant in terms of nuclear processes determined in this work. However, these various normally occurring nuclides do (based on their occurring by their natural relative abundances) demonstrate that the SIMS is measuring single nuclides and not clusters. Moreover, the nuclides that show anomalous mass intensities are emphasized here as proof of nuclear transmutations occurring in the magnet. Therefore the anomalous $^{57}$Fe, $^{51}$V, $^{28}$Si, $^{28}$Na, $^{37}$Cl, $^{72}$Ge $^{50}$Ti, $^{48}$Ti , $^2$H,

$^3$H and $^4$He mass intensities are real effects not artifacts associated with clustering of lighter elements during the mass spectroscopy. These elements exhibiting anomalies are culprits to the predicted magnetically enhanced pycnonuclear reactions and the associated spin induced orbital phenomena.

The SIMS reveals the unusual presence of lighter 4d nuclides than Ag like $^{89}$Y, $^{90}$Zr, $^{91}$Zr, $^{92}$Zr, $^{92}$Mo and $^{93}$Nb after the prolong use of the coils in the magnet. The unused Cu-Ag coil lacks these nuclides of masses 82, 84, 86, 90, 91, 92, and 98 Da. See Figure 2. After prolong use of the Cu-Ag coil in the magnet all masses between 80 and 100 are observed with the exception of mass 89. See Figure 4. Mass 89 corresponds exactly with the element Yttrium. It appears that the transmutations during the prolong operation of the magnet transmutes $^{89}$Y to heavier nuclides. $^{89}$Y disappears from the coil after prolong use. Heavier nuclides than $^{89}$Y like $^{90}$Zr, $^{91}$Zr, $^{92}$Zr, $^{92}$Mo and $^{93}$Nb form from the $^{89}$Y. Of the predominant elements of lower masses ($^{63}$Cu, $^{39}$K, $^{40}$Ca, $^{27}$Al, $^{28}$Si, $^{23}$Na, $^{16}$O, and $^{12}$C), neither $^{63}$Cu$^{27}$Al (90 Da) nor $^{63}$Cu$^{28}$Si (91 Da) clustering account for the masses 90 and 91 Da on the basis of such clusters being inconsistent with the relative abundances of $^{63}$Cu, $^{27}$Al and $^{28}$Si nuclides. The mass intensity of $^{91}$Zr is too high relative to $^{90}$Zr. Zr is only element having isotopes of masses 90 and 91 Da. The SIMS intensities of the 90 and 91 Da peaks do not correspond with the natural relative abundances of $^{90}$Zr and $^{91}$Zr. These Zr nuclides are therefore formed by transmutation during the operation of the magnet. The masses of 92 and 93 Da are also not consistent with the relative abundances of $^{65}$Cu and $^{27}$Al and $^{28}$Si. Zr is the only element with a nuclide of mass 92 Da. $^{92}$Zr is too high for it to correspond to the relative abundance with $^{90}$Zr and $^{91}$Zr. There is only one element with mass 93 that is Nb. The $^{92}$Zr and $^{93}$Nb are also formed by transmutation processes during the operation of the magnet. On the basis of these discrepancies in natural relative abundances of possibly Cu, Al, and Si clusters, it appears that the nuclides of masses 90, 91, 92, and 93 Da are formed by some transmutations processes during the operation of the magnet. The observed disappearance of $^{89}$Y in the used coils is consistent with the appearance of these heavier nuclides of masses 90, 91, 92, and 93 Da. See Figures 2 and 4. It is harder to compare nuclides of masses 94-100 because they are common masses of nuclides of different elements. With the increasing importance of clustering beyond Cu, it is possible that the masses 95-97 Da may correspond to $CuO_2$. The appearance of mass 98 Da after use of the Cu-Ag coils is also evidence of anomalous nuclear transmutations. Mass 98 Da corresponds to either to $^{98}$Mo, $^{98}$Tc or $^{98}$Ru. $^{98}$Tc is very radioactive. See Figure 4. The transmutation reactions occurring to form these nuclides of masses 90, 91, 92, 93 and 98 Da may involve the electron capture of Ag nuclides with neutron loss to form successively lighter nuclides of smaller atomic numbers.

The SIMS data also reveal the unusual presence of Ru, Rh, and Pd nuclides in the used Cu-Ag coils. The SIMS also demonstrated anomalous nuclide ratios of these elements in the used Cu-Ag coils. See Figures 3 and 5. Prior to use of the coils $^{100}$Mo, $^{100}$Ru, $^{102}$Ru, $^{104}$Ru and $^{104}$Pd seem to be present. See Figure 3. These masses of 102, and 104 Da cannot be CuK, CuAr or CuCa clusters. See Figure 2. Some of the $^{23}$Na, $^{24}$Mg, $^{40}$Ca and $^{39}$K come from impurities in the Cs beam of the CIMS. Some of these nuclides are also formed by transmutation reactions in the coil during the operation of the magnet. $^{63}$Cu and $^{65}$Cu nuclides exist in the Cu-Ag coil. On the basis of the mass spectra not much CuNa (86 Da and 88 Da) forms during SIMS this is evidence of limited clustering. See Figure 2. The observed SIMS peaks at 102 and 104 Da are inconsistent with CuK, CuAr or CuCa clusters forming, because the relative ratios of the 102, 104 and 105 Da peaks are not consistent with the natural relative abundances of Cu, K, Ar and Ca isotopes. K is mostly abundant as $^{39}$K. Ar is present mostly as $^{40}$Ar. Ca is present mostly as $^{40}$Ca. Clustering of Cu and K or Ar or Ca is also not suspected because no 103 Da peak ($^{63}$Cu$^{40}$Ca or $^{63}$Cu$^{40}$Ar) is observed in Figures 3 and 5. If these lighter Cu, Ar, K and Ca nuclides clustered to form mass signals at 102, 104 and 105 Da, then they should also produce a signal at 103 Da in Figure 3. But no signal at 103 Da is observed in Figure 3 so the clustering is not occurring during SIMS. In the initial unused Cu-Ag coil, masses 105, 107 and 109Da also appear. The mass 105 Da is likely $^{105}$Pd. The masses 107 and 109 Da are $^{107}$Ag and $^{109}$Ag nuclides. It is important to note the absence of masses 101, 103 and 106 Da in the coil prior to use in the magnet. See Figure 3. These missing masses correspond to $^{101}$Ru, $^{103}$Rh, and $^{106}$Pd nuclides. After the use of the coil the mass spectra drastically change in the mass range of 100- 109 Da. See Figure 5. Masses of 102 and 104 Da dramatically increase, but the mass 105 Da does not proportionately increase with masses 102 and 104 Da, which eliminates the possibility of clustering by (CuK, CuCa, CuAr) for the explanations of these masses. The natural relative abundances of Cu, Ca , K and Ar do not account for the observed relative SIMS intensities of the 102 and 104 Da peaks. However, the relative abundances of $^{102}$Ru and $^{104}$Ru agree with this SIMS data at 102 and 104 Da in Figure 5, but the $^{101}$Ru mass intensity does not correspond to the relative abundances of $^{102}$Ru and $^{104}$Ru. After the use of the coil, the new peaks at 101, 103, and 106 Da appear. See Figures 3 and 5. The relative abundance of $^{106}$Pd and $^{108}$Pd should be about equal, but the data here determines $^{106}$Pd $\gg$ $^{108}$Pd. These masses of 101, 102, and 106 Da correspond to the formation of $^{101}$Ru, $^{103}$Rh, $^{106}$Pd, and $^{106}$Cd during the prolong use of the Cu-Ag coil. The formations of $^{101}$Ru, $^{103}$Rh, and $^{106}$Pd nuclides are consistent with them forming from Ag, since the relative amounts by SIMS intensities are such that $^{101}$Ru $<$ $^{103}$Rh $<$ $^{106}$Pd. The electron capture by Ag nuclei with neutron release results in the formation of these elements of smaller atomic number and mass number in the coils during the operation of the magnet. The electron capture with the release of neutrons by the excited nuclides results in the Ag forming Pd; the Pd can by an e$^-$ capture and the neutron release process form Rh; the Rh can by an e$^-$ capture and neutron release process can form Ru. These e$^-$ capture cross-section processes should correlate with spin moments of the various nuclides on the basis of moments increasing: Ag $<$ Pd $<$ Rh $<$ Ru. These nuclides $^{101}$Ru, $^{103}$Rh, and $^{106}$Pd were observed by both SIMS and ICP-MS (Table 5) in the used coils. These nuclides $^{101}$Ru, $^{103}$Rh, and $^{106}$Pd were further demonstrated not to be CuK or CuCa by use of O$_2^+$ beam rather than the Cs$^+$ beam, the O$_2^+$ beam lacks Na, Mg and K, Ca impurities. When using the O$_2^+$ beam the masses between 100 and 110 were still observed, eliminating the possibility that these masses are clusters of CuK or CuCa. The observation of these elements $^{101}$Ru, $^{103}$Rh, and $^{106}$Pd is unusual because of their low abundance on the surface of the earth. ***The SIMS revealed unusual isotopic ratios of these various rarely occurring nuclides.*** These rare nuclides are formed by transmutations in the Cu-Ag coil during the prolong operation of the magnet.

The SIMS data also exhibit unusually high mass intensities associated with Ag fragmenting into various lighter nuclides like $^{85}$Rb and $^{23}$Na; $^{57}$Fe and $^{50}$Ti; $^{55}$Mn and $^{52}$Cr; $^{55}$Mn and $^{51}$V; $^{58}$Ni and $^{51}$V; $^{68}$Zn and $^{41}$K; $^{79}$Br and $^{28}$Si; $^{37}$Cl and $^{72}$Ge. The unusual SIMS intensities relative to the naturally occurring abundances of these various transmutation products determine strong evidence of this fission of Ag occurring in the Cu-Ag coil during the operation of the magnet. Although these other nuclides of other elements (Li, Be, B, C, O, F, Ne, Mg, P, Sc) within the mass range of these anomalous nuclides do appear according to their natural relative abundances, these anomalously observed elements $^{85}$Rb, $^{79}$Br, $^{72}$Ge, $^{68}$Zn, $^{58}$Ni, $^{57}$Fe, $^{55}$Mn, $^{51}$V, $^{52}$Cr, $^{50}$Ti, $^{41}$K, $^{37}$Cl and $^{23}$Na) exhibit unusual mass spectra, which is very interesting. Moreover, the fact that these few nuclides with unusual mass spectra intensities may be grouped together such that their masses to sum to the mass of either $^{107}$Ag or $^{109}$Ag is strong evidence of them originating by this low temperature fission from Ag. $^{107}$Ag is fragmented into $^{57}$Fe and $^{50}$Ti or. $^{50}$Ti and $^{57}$Fe reveal unusually higher mass peaks in the Cu-Ag coil after use of the coil in the magnet. Above we already demonstrated that lighter elements cannot account for the unusual observed masses of $^{57}$Fe and $^{50}$Ti. $^{109}$Ag may be forming $^{57}$Fe and $^{52}$Cr. $^{52}$Cr is also exhibiting an unusually high mass intensity relative to other nuclides of Cr. $^{107}$Ag may also be fragmenting to form $^{55}$Mn and $^{52}$Cr or $^{51}$V. The mass intensities of $^{55}$Mn and $^{51}$V are unusually higher than their corresponding isotopes in the used Cu-Ag coils. $^{51}$V may also be considered to form from $^{109}$Ag fragmenting into $^{58}$Ni and $^{51}$V. The mass intensity of $^{58}$Ni in the used Cu-Ag coils exhibits unusually higher mass relative to the intensity of $^{60}$Ni. The lighter elements, which have been demonstrated to be observed as single nuclides do not appear based on the natural abundances to cluster to form these anomalous mass intensities of $^{55}$Mn, $^{52}$Cr, $^{51}$V, and $^{58}$Ni masses. The mass spectra of the Cu-Ag coil after use in the magnet also determines that $^{109}$Ag is fragmenting into $^{68}$Zn and $^{41}$K. The mass intensity of $^{41}$K is unusually high. The mass intensity of $^{68}$Zn is also unusually high. The SIMS data of the used Cu-Ag coil are also consistent with $^{107}$Ag forming $^{79}$Br and $^{28}$Si. $^{79}$Br mass intensity is unusually high relative to other isotopes of Br. The mass intensity of $^{28}$Si is unusually high relative to $^{28}$Si and $^{29}$Si. The $^{79}$Br and $^{81}$Br SIMS peaks are not consistent with their natural relative abundances. These anomalous masses of Br isotopes cannot be accounted for by CuO clustering, because the relative SIMS intensities of the $^{79}$Br and $^{81}$Br peaks are inconsistent with the natural relative abundances of Cu and O so these cannot account for the anomalous $^{79}$Br peak. $^{109}$Ag fragmentation into $^{85}$Rb and $^{23}$Na is also consistent with the mass spectra of the used Cu-Ag coils. $^{85}$Rb is too high relative to $^{87}$Rb. CuNa clusters cannot account for the anomalous intensity of $^{85}$Rb on the basis of inconsistent SIMS intensities with the relative abundances of Cu and Na. $^{85}$Rb forms from $^{109}$Ag fragmenting in coil. The SIMS intensities are consistent with $^{85}$Rb absorbing neutron and fragmenting into $^{63}$Cu and $^{23}$Na. The $^{63}$Cu is unusually higher than the mass intensity of $^{65}$Cu in the used Cu-Ag coil. $^{37}$Cl and $^{72}$Ge exhibit anomalous mass intensities in the used Cu-Ag coil. The Ag is fragmenting into $^{37}$Cl and $^{62}$Ge during the operation of the magnet.

The fact that these various nuclides ($^{85}$Rb, $^{79}$Br, $^{72}$Ge, $^{68}$Zn, $^{58}$Ni, $^{57}$Fe, $^{55}$Mn, $^{50}$Ti, $^{52}$Cr, $^{51}$V, $^{39}$K, $^{37}$Cl, $^{28}$Si, and $^{23}$Na) all exhibit (in a nonrandom manner) anomalous mass intensities in the used Cu-Ag coil; their masses cannot be accounted for by the clustering of smaller nuclides; the masses of all neighboring nuclides of similar masses (to these anomalously observed nuclides) are consistent with their natural relative abundances; all these anomalously observed nuclide masses may be grouped such that they sum to isotopes of Ag; and the Little theory that magnetization of the huge current density in the Cu-Ag coil under strong pressure and thermal stresses cause pycnonuclear reactions, all these facts demonstrate that these mass anomalies result from the fission of Ag during the use of the coil and the theory of magnetically enhanced pycnonuclear phenomena is real. All of the anomalies in the mass spectra can be explained by this Little theory that pycnonuclear reactions occur and are explained by spin induced torque of electrons and protons within the Cu-Ag lattice for various reverse beta, neutron capture, electron capture and proton capture processes. The formation of these elements in the coil during the operation of the magnet contributes to the sudden increase in the coil resistance during the lifetime of the coil. Prior ideas assumed that the increased resistance resulted from the erosion of Ag during the operation of the magnet. However, the drastic increase in the coil resistance cannot be fully accounted for by the simple loss of Ag. As Ag is loss, the Cu-Ag coil still consists of Cu so its resistance should still be low. Actually the original Cu-Ag unused coil has higher resistance than a pure Cu coil. So while the Ag is being dissolved from the coil, the coil's resistance should drop as the coil becomes more pure Cu. But this is not what is observed. Instead, as Ag is dissolved, the resistance of the coil increases. Here we suggest that this increase in resistance is a result of impurities formed by the transmutation of the Ag in addition to the dissolution of the Ag. It is important to note that the sudden rise in resistance of the coil is accompanied by a sudden rise in its temperature. As the transmutations occur, the coil should get hotter. The transmutations generate elements like Br, Si, Ge, K, Na and Cl, which should cause the resistance to increase. So both the change in resistance and temperature of the coil are explained by this theory and data of magnetically enhanced pycnonuclear reactions. However, on the basis of this theory and mass data for magnetically enhanced pycnonuclear reactions during the operation of the magnet, the transmutations of Ag and the formations of elements like Rb, Br, Ge, K, Si and Na would lead to a release of excess heat and to a dramatic increase in the resistance of the Cu-Ag coil.

The SIMS data also revealed unusual nuclides of masses in the range of 110-119 Da in the used coils. See Figures 3 and 5. These heavier nuclides beyond Ag were observed in the coils by SIMS after the strong magnetization during prolong use of the Cu-Ag coils in the magnet. Nuclides of masses 112 and 115 Da are observed in the unused coils. These nuclides correspond to $^{112}$Cd or $^{112}$Sn and $^{115}$In or $^{115}$Sn. The observed disappearance of $^{112}$Cd in the used coils is consistent with the appearance of $^{113}$In. See Figures 3 and 5. In Figure 5, it is important to note that the $^{110}$Cd and $^{111}$Cd appear by their natural relative abundances in the SIMS, but $^{112}$Cd is missing. The $^{115}$In remains after prolong use of the coil. A nuclide of mass 117 Da is observed in the coil before its use. After use of the coil, the nuclide of mass 117 Da disappears. This nuclide of mass 117 Da is $^{117}$Sn. During the operation of the magnet, the $^{117}$Sn is observed to transform to masses of 118, 119, 120, 121, 122, and 123 Da. It may be argued that these new peaks are $^{63}$Cu$^{58}$Ni and $^{63}$Cu$^{60}$Ni. But the observed relative SIMS intensities at these masses are inconsistent with the natural relative abundances of these lighter Cu and Ni isotopes. The relative abundances of Cu and Ni also do not account for the mass intensities of 119, 120 and 121 Da. These masses correspond to $^{119}$Sn, $^{120}$Sn, $^{121}$Sb and $^{123}$Sb. Figures 3 and 5. The formation of these Sn and Sb nuclides from Ag is consistent with the various thermal neutron absorption cross-sections of nuclides from Ag to Cd to In to Sn to Sb. Thermal neutrons formed by the reverse beta during operation of the magnet transform Ag to Cd to In to Sn and to Sb. The beta process of these resulting neutrons rich nuclides can lead to increase atomic number and other elements like Cd, In, Sn, and Sb

*All of these different observations of anomalous abundances and relative amounts of $^2$H, $^3$H, $^3$He, $^4$He, $^{23}$Na, $^{28}$Si, $^{37}$Cl, $^{41}$K, $^{50}$Ti, $^{52}$Cr, $^{51}$V, $^{55}$Mn, $^{57}$Fe, $^{58}$Ni, $^{63}$Cu, $^{68}$Zn, $^{72}$Ge, $^{79}$Br, $^{85}$Rb, $^{89}$Y, $^{90,91,92}$Zr, $^{92}$Mo, $^{93}$Nb, $^{101}$Ru, $^{103}$Rh, $^{106}$Pd, $^{112}$Cd, $^{115}$In, $^{119,120}$Sn and $^{121,123}$Sb nuclides by different mass analytic techniques (SIMS and ICP-MS) provide strong evidence of enhanced pycnonuclear phenomena occurring within the Cu-Ag coil under the strong magnetization (45 tesla), large electric potential (403 volts), large electric current (74,000 amps), huge Lorentz compression, and thermal fluctuations over prolong operation times (>2000 hours).*

**Conclusion**
Just as the external magnetic field orients spins within the paramagnetic Cu-Ag coils and the solvated protons within the DC magnets for novel magnetocatalytic chemistry, similar effects occur within paramagnetic iron beyond its Curie temperature such that high spin orientation by strong external magnetic field induced orbital rehybrization, stabilization and orientation of carbon radicals for diamond nucleation and growth on the basis of this Little Effect [26,27]. Within this system high spin---spin interactions between carbon atoms and metal atoms within the liquid growth interface cause orbital rehybridization of carbon from sp to $sp^2$ and to $sp^3$ orbital motions for incorporation within the diamond edge. These magnetocatalytic spin induced orbital dynamics under the intense thermal conditions according to the Little Effect change orbital symmetry of carbon contrary to the Woodward-Hoffmann Rule [41]. But the total orbital and spin dynamics of both the carbon and the catalysts obey the Woodward-Hoffman Rule [41]. In both the Cu-Ag-water and the Fe-C systems the spin – spin interactions under greater thermal conditions induce the orbital dynamics for novel electron transfer and bond rearrangements on the basis of this Little Effect. The Little Effect also provides greater insight and understanding of other systems. The Little Effect provides greater understanding of the ferromagnetism of Fe, Co, and Ni [42] on the basis of spin induced orbital dynamics and hybrid 3d and 4s orbitals for greater coulomb energy relative to exchange energy of bonding and greater correlated energy associated with polarizing localized electrons. On the basis of the Little Effect similar spin induced interactions are weaker for 4d transition metals so such metals are not ferromagnetic. The Little Effect provides explanations to the ferromagnetism of transition metal borides, carbides, nitrides and oxides [43]. The recent nonconventional magnetism within boronic, carbonaceous, and nitrogenous materials [44] is accounted for by the Little Effect. The Little Effect provides an explanation of high temperature superconductivity [45] on the basis of phonons exciting (Kasha Effect) [9] conduction electron pairs into high spin states by El-Sayed Effect [8] with the high spin states maintaining the pairing and inducing the needed rehybridization (Little Effect) to the ground conduction state. The Little Effect determines the chemistry of high temperature superconductivity and also demonstrates and integrates the roles of diamagnetism, ferromagnetism and ferrimagnetism to superconductivity in a way not done before. The Little Effect offers greater understanding of controversial phenomena associated with large thermal and sporadic effects of aqueous palladium electrochemistry [46]. On the basis of the Little Effect pycnonuclear phenomena are explained and enhanced. The Little Effect predicts that someday strong external magnetic field will allow the spin polarization of electrons and protons within the palladium lattice for the accumulation of sufficient electromagnetic energy to drive reverse beta processes for more reproducible and control of cold fusion phenomena.

**Acknowledgement:**

With external gratitude to GOD.

For my three sons: Reginald Bernard, Ryan Arthur and Christopher Michael.

* Corresponding author: redge_little@yahoo.com

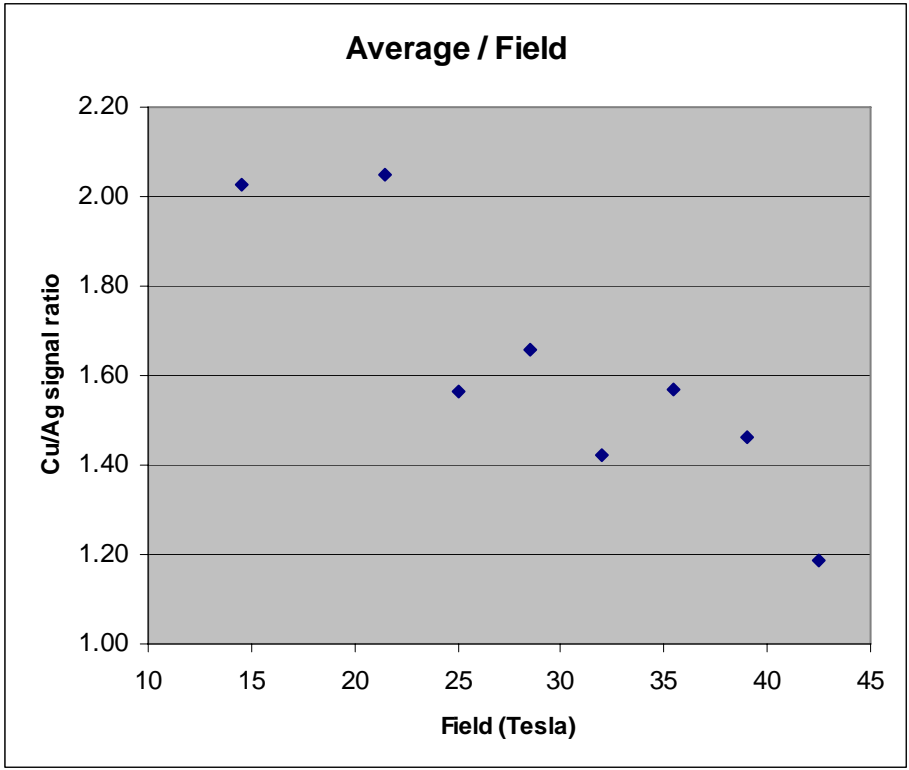

**Figure 1**

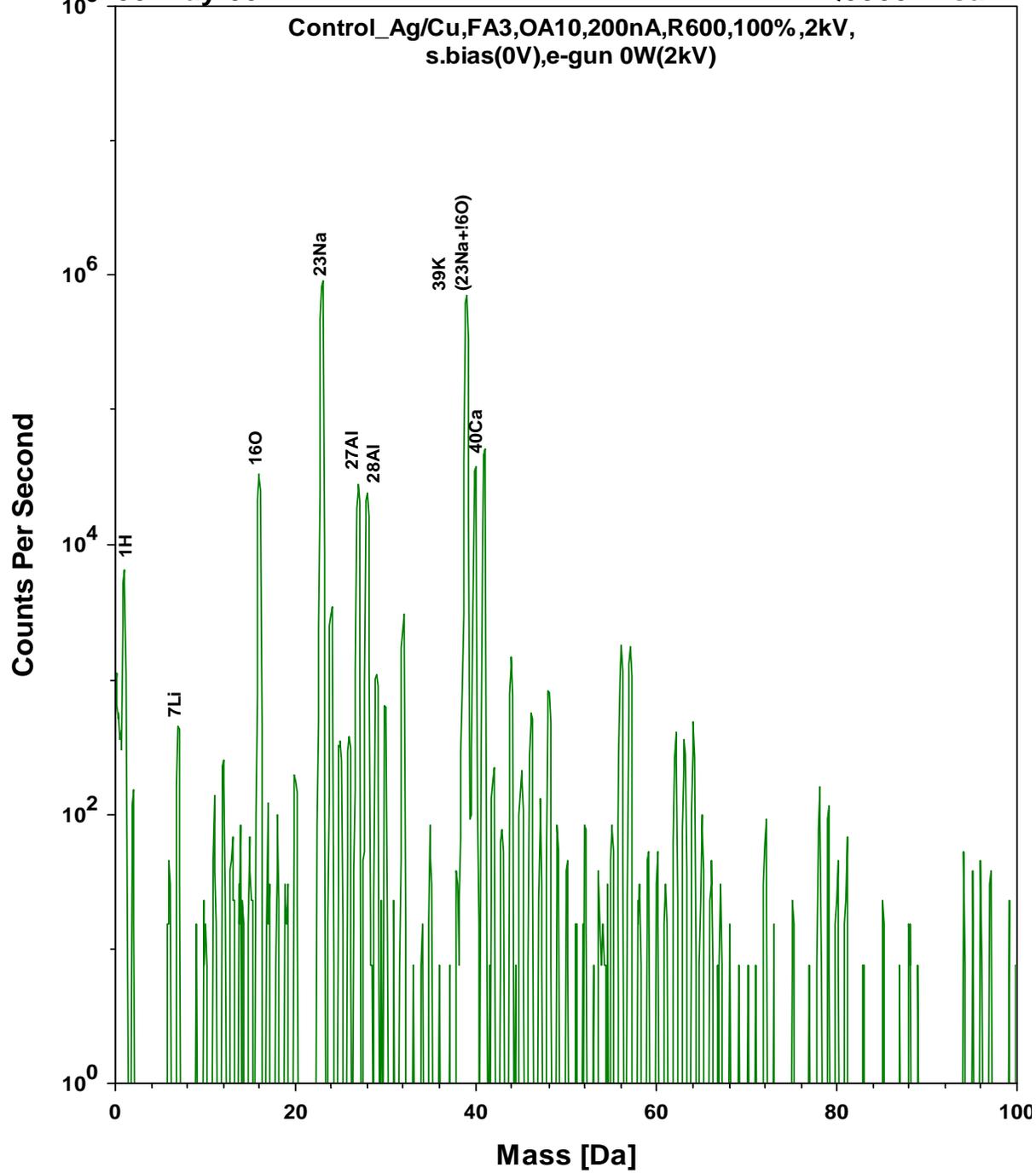

**Figure 2**

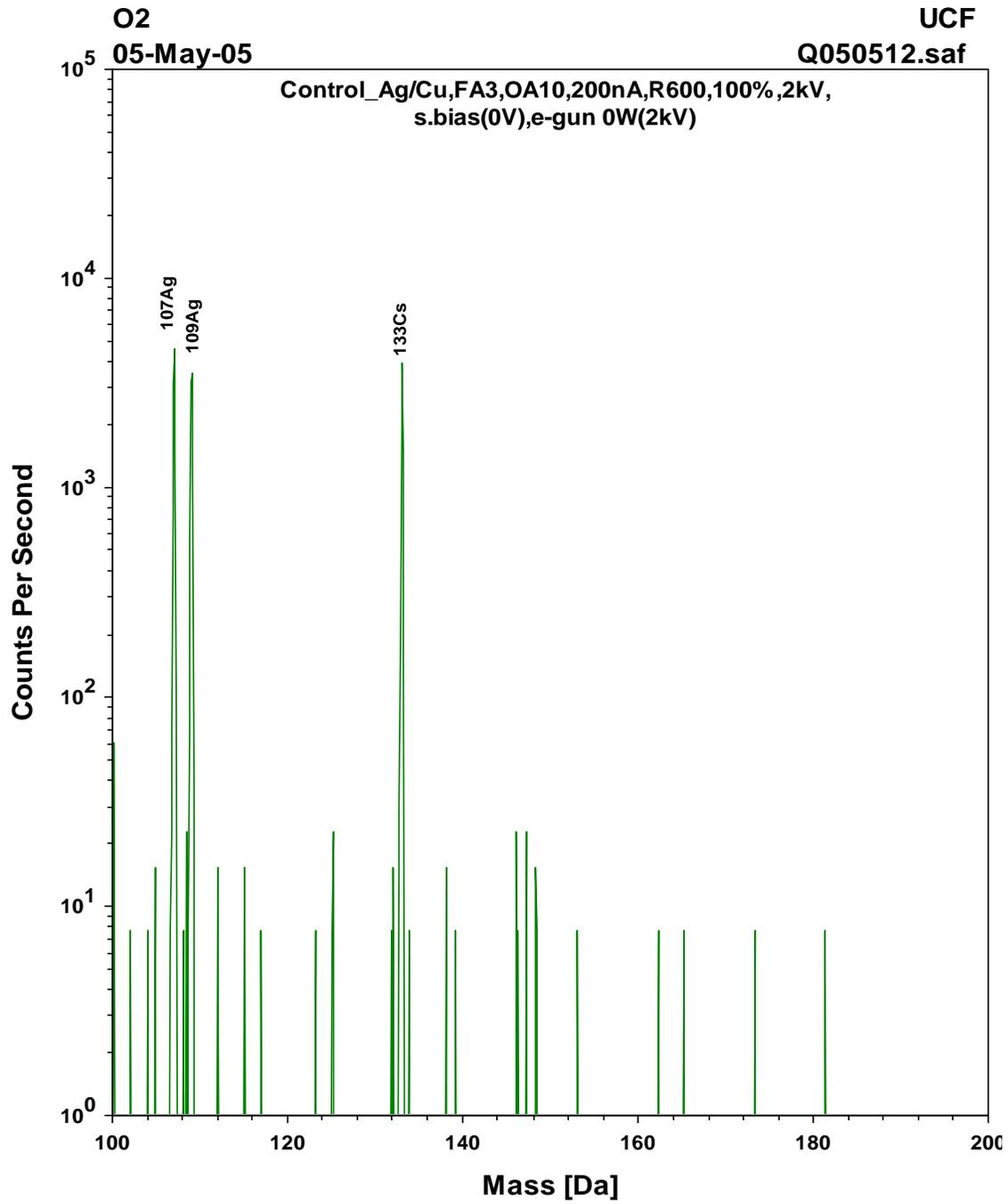

**Figure 3**

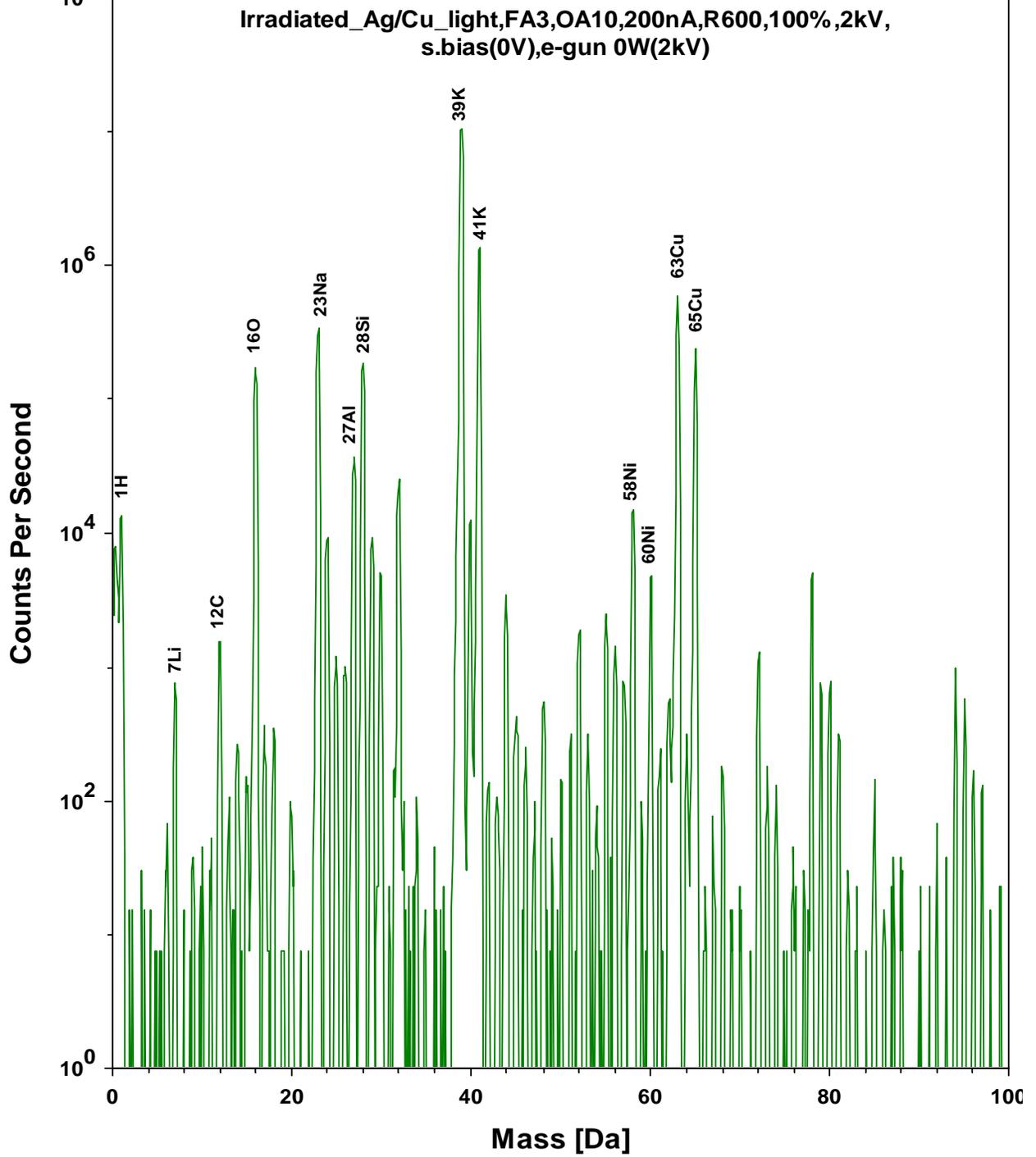

**Figure 4**

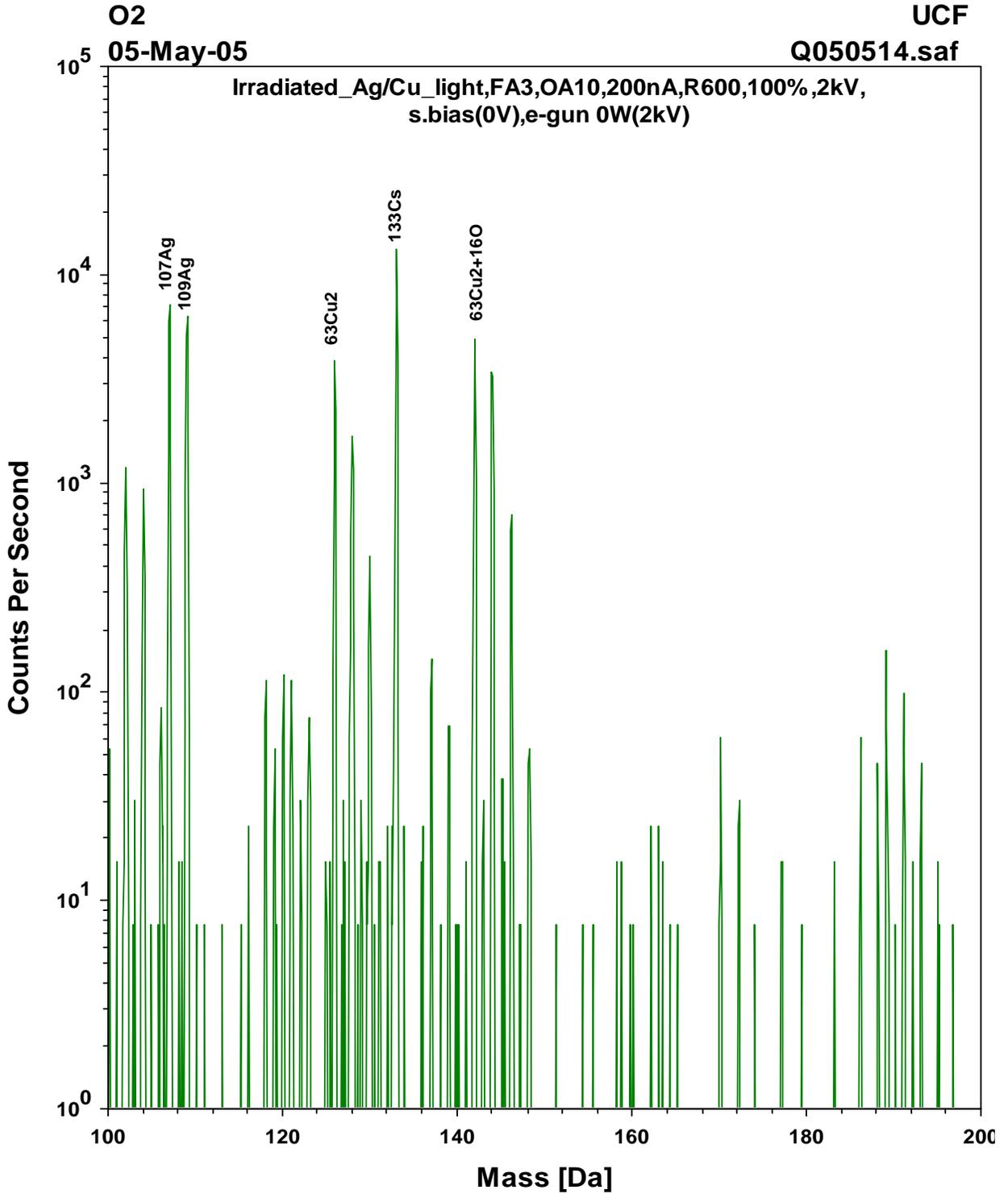

**Figure 5**